\documentstyle[12pt,epsfig,axodraw]{article}
\newcommand{\pslash}{\not{\!P}}
\newcommand{\pnslash}{\not{\!P_N}}
\newcommand{\snslash}{\not{\!S_N}}
\newcommand{\xislash}{\not{\!\xi}}

\newcommand{\cslash}{\not{\!\xi}}

\newcommand{\be}{\begin{equation}}
\newcommand{\ee}{\end{equation}}
\newcommand{\ba}{\begin{eqnarray}}
\newcommand{\ea}{\end{eqnarray}}

\newcommand{\nsigma}{\mbox{\boldmath $\sigma$}}
\newcommand{\ncalR}{\mbox{\boldmath ${\cal R}$}}

\newcommand{\nOmega}{\mbox{\boldmath $\Omega$}}

\newcommand{\nchi}{\mbox{\boldmath $\chi$}}

\newcommand{\nl}{{\bf      l}}
\newcommand{\nn}{{\bf      n}}

\newcommand{\nk}{{\bf      k}}
\newcommand{\np}{{\bf      p}}       
\newcommand{\nq}{{\bf      q}}
\newcommand{\nr}{{\bf      r}}         
\newcommand{\ns}{{\bf      s}}

\newcommand{\nx}{{\bf      x}}
\newcommand{\ny}{{\bf      y}}
\newcommand{\nz}{{\bf      z}}

\newcommand{\nM}{{\bf      M}}

\begin{document}
\begin{titlepage}
\mbox{} 
\vspace*{2.5\fill} 
{\Large\bf 
\begin{center}

Skewed recoil polarization in $(e,e'p)$ reactions
 from polarized nuclei

\end{center}
}
\vspace{1\fill} 
\begin{center}
{\large 
J.E. Amaro$^{1}$,
M.B. Barbaro$^{2}$,
J.A. Caballero$^{3}$
}
\end{center}
\begin{small}
\begin{center}
$^{1}${\em 
          Departamento de F\'{\i}sica Moderna, 
          Universidad de Granada,
          Granada 18071, Spain}  \\[2mm]
$^{2}${\em 
  Dipartimento di Fisica Teorica,
  Universit\`a di Torino and
  INFN, Sezione di Torino \\
  Via P. Giuria 1, 10125 Torino, ITALY} \\[2mm]
$^{3}$ {\em 
          Departamento de F\'\i sica At\'omica, Molecular y Nuclear \\ 
          Universidad de Sevilla, Apdo. 1065, E-41080 Sevilla, SPAIN
}
\end{center}
\end{small}

\kern 1. cm \hrule \kern 3mm 

\begin{small}
\noindent
{\bf Abstract} 
\vspace{3mm} 

\noindent
The general formalism describing $\vec{A}(\vec{e},e'\vec{p})B$
reactions, involving polarization of the electron beam, target and
ejected proton, is presented within the framework of the relativistic
plane wave impulse approximation for medium and heavy nuclei.  It is
shown that the simultaneous measurement of the target and ejected
proton polarization can provide new information which is not contained
in the separate analysis of the $\vec{A}(\vec{e},e'p)B$ and
$A(\vec{e},e'\vec{p})B$ reactions.  The polarization transfer
mechanism in which the electron interacts with the initial
nucleon carrying the target polarization, making the proton exit
with a fractional polarization in a different direction, is referred
to here as ``skewed polarization''.  The new observables
characterizing the process are identified, and written in terms of
polarized response functions and asymmetries which are of tensor
nature.  The corresponding half-off-shell single-nucleon responses are
analyzed using different prescriptions for the electromagnetic vertex
and for different kinematics.  
Numerical predictions are presented for selected perpendicular and
parallel kinematics in the case of $^{39}$K as polarized target.

\kern 2mm 

\noindent
{\em PACS:}\ 
25.30.Fj,  
24.10.Jv,  
24.70.+s   

\noindent
{\em Keywords:}\ Nuclear reactions; Coincidence electron scattering;
Polarization transfer;
Response functions; Polarized momentum distribution.

\end{small}

\kern 2mm \hrule \kern 1cm
\end{titlepage}


\section{Introduction}


It is well known that polarization degrees of freedom in electron scattering
reactions lead to new observables which provide additional information on the
nuclear structure as well as on the reaction mechanism
\cite{Frul85,RaDo89,Kel96,Bof96}. 
In general, the control of leptonic and hadronic polarizations allows new combinations of
  electromagnetic multipole matrix elements that show different sensitivities
  to the diverse ingredients entering in the description of the reaction
  mechanism.  

In inclusive $(\vec e,e')$ processes, the polarization of the incident
lepton yields information not only on the nucleonic structure (in particular 
the strange and axial nucleon's form factors, see e.g. 
refs.~\cite{Donnelly:1991qy,Musolf:1993tb}), but also on some specific nuclear 
correlations which are not probed by unpolarized 
electrons~\cite{Alberico:1993ur,Barbaro:1993jg,Barbaro:1995ez,Barbaro:1995gp}.

In recent years a great effort, both from experimental and
  theoretical points of view, has been devoted to the analysis of 
  exclusive $(e,e'p)$
  with leptonic and/or hadronic polarization measurements for medium nuclei.
  This has been for instance the case of $(e,e'\vec{p})$ processes and the
  measurement of induced polarization \cite{Woo98,TJNAF} as it crucially
  depends on final state interactions (FSI) (the induced polarization is zero
  in the plane wave limit) \cite{Udi00,Ito97,Ryc99,Kel99,Kel99b,Joh99,Kaz04}.
  Double polarized
  $(\vec{e},e'\vec{p})$ experiments have been also carried out to
  measure polarization transfer asymmetries for medium \cite{Mal99,Mal00} and
  light nuclei \cite{Die01,Strauch03}, as they may provide information on the
  possible modifications of the nucleon form factors in the nuclear interior
  \cite{Ryc99,Cris04}.  Finally, the case of electron scattering on polarized
  targets, i.e., $\vec{A}(e,e'p)$ and $\vec{A}(\vec{e},e'p)$ processes, has
  been analyzed as well from the theoretical point of view for medium
    nuclei in a number of
  papers~\cite{Boffi88,Cab93,Cab94,Edu95,JAC95,Ama98b} and, in particular,
    important connections between FSI and the nuclear polarization direction
  have been found \cite{Ama99,Ama02}.
  
  However studies of
  spin observables in $\vec{A}(\vec{e},e'\vec{p})$ when the beam, target and
  final nucleon are simultaneously polarized exist only scarcely. Apart from
  the general approach of ref. \cite{RaDo89}, research on these reactions has
  only been performed
  for light nuclei. In particular it is worth mentioning
  the systematic research on general sets of spin observables for deuterium
  performed in the last decade by Arenh\"ovel {\em et al.}
  \cite{Are04,Are02,Are00,Are98,Are95,Are92}. In the present
  paper we try to fill
  this gap and begin an exploration of the new spin observables that arise in
  these exclusive reactions from medium nuclei. 
  Although no experiments of this kind are presently under planning,
  theoretical studies of such reactions are
  needed to gauge the magnitude and properties of the new observables, and to
  determine if new relevant physical information can be extracted from them.
  Accordingly, in this work we focus on a new polarization mechanism of the
  recoil proton, which is absent in the reactions previously studied. In fact,
  for unpolarized nuclei, there are two such mechanisms: (i) For unpolarized
  electrons {\em induced polarization} is produced on the recoil protons by
  effect of the FSI with the residual nucleus.  In DWIA models \cite{Ama04}
  this effect is partially due to the imaginary part of the central optical
  potential ---which produces different absorptions for protons exiting from
  regions of the nucleus with different local spin directions--- and partially
  due to the spin-orbit potential, which produces a spin-dependent
  interaction.  (ii) Polarized electrons can transfer part of their spin to the
  nucleon. Thus {\em polarization transfer} arises already within the plane wave limit and
  FSI are proved not to affect appreciably these observables for low missing
  momentum \cite{Cris04,Kaz03}.  In the case of unpolarized electrons on polarized
  nuclei another mechanism takes place, namely the polarization transfer from
  the target to the final proton.  In some cases of interest, a semi-classical
  interpretation of the reaction mechanism producing such polarization
  transfer can be drawn.  In fact, in the case of knock-out from the outer
  shell, the initial proton is carrying part of the nuclear polarization, so
  an expected value of the initial spin can be defined for the proton
  \cite{Ama02,Ama04}.  In general, the final direction of the recoil
  polarization does not coincide with the initial one, so the interaction is
  producing a change of the proton spin.  The
  effect of the electromagnetic interaction is then to pull the proton with a
  twist of its spin, so the final proton results partially
  polarized in a direction different from the initial one.  We call this effect
  ``skewed polarization'', to distinguish it from the term
  ``polarization transfer'' normally used for polarized electrons. The
  corresponding hadronic observables, namely response functions and recoil
  polarization asymmetries, can be represented by tensor quantities.  One of
  the goals of this paper is to present predictions for the
  polarization asymmetries that can be
  measured in these reactions for different polarization directions and to
  identify the situations in which large recoil polarizations are expected in
  a typical experiment.

It is well known that
FSI constitute an essential ingredient in order to analyze experimental data
(cross sections and/or response functions) in detail.  Some specific
polarized and unpolarized responses are strictly zero in the plane wave limit;
hence these observables may help to constrain theoretical uncertainties in
the treatment of FSI.  However, although being aware of the importance of FSI
or other effects due to ingredients beyond the impulse approximation such as
meson exchange currents (MEC), in this work we restrict our analysis to the
plane wave impulse approximation (PWIA) in a first 
approach to the problem. As already noted  in
refs.~\cite{Cab93,Cab98a,Cab98b,Cris02}, this particularly simple
description allows one to simplify and clarify the essential physical issues
underlying the problem.  Moreover, some observables given as ratios of cross
sections and/or response functions are shown to be rather insensitive to FSI
and other distortion effects, and thus PWIA calculations may be adequate.

In spite of its simplicity, PWIA has proved its capability for the
interpretation of experimental data. One of the basic advantages of PWIA is
that the cross sections, and consequently the response functions, factorize
into a single-nucleon cross section, dealing directly with the interaction
between the incident electrons and the bound nucleons inside the nucleus, and
a spectral function which gives the probability to find a nucleon in the
nucleus with given energy and momentum. Both terms may 
in turn be functions of the spin variables involved in
the reaction.  The importance of factorization lies on the fact that
experimental data are still usually interpreted making use of an 
{\em effective}
 spectral function that is extracted from experiment. A systematic
study of the validity of factorization within the general framework of the
relativistic distorted wave impulse approximation (RDWIA) has been presented
recently in~\cite{Vign04}.  It has been shown that exact factorization also emerges
with FSI under very restrictive assumptions on the relativistic dynamics and
spin dependence in the problem. For some observables, evaluated as ratios of
cross sections, and not too high values of the missing momentum, it has been
proved that the {\em factorized} PWIA results are in accordance with much more
sophisticated and {\em unfactorized} calculations which include FSI.
These ideas apply to the skewed polarization asymmetries 
analyzed in this paper, since these observables arise already in PWIA: 
hence one expects the effect of 
FSI to be small on these, at least for low values of the missing momentum.

In this paper we follow closely the formalism
developed in~\cite{Cab93,Edu95} for the case of $\vec{A}(\vec{e},e'p)$, i.e.,
scattering on polarized targets, and in~\cite{Cris02} for outgoing nucleon
polarization reactions $A(\vec{e},e'\vec{p})$. Here we extend the analysis and
consider the case of target and outgoing nucleon polarizations measured
simultaneously.  Obviously, the single nucleon responses entering into the
analysis of $\vec{A}(\vec{e},e'\vec{p})$ reactions contain as particular cases
those ones entering into the separate descriptions of $A(\vec{e},e'\vec{p})$
and $\vec{A}(\vec{e},e'p)$ processes. Therefore
we focus on the ``new'' response functions, which depend on both the
target and ejected nucleon polarizations.

A basic ingredient of PWIA deals with the single-nucleon
response functions and how they depend on the off-shell
effects for different kinematics. In this paper we    
perform a study of these effects on the new response functions
 following the approach originally introduced
by de Forest~\cite{For83}. We make use of the two standard forms for
the $\gamma NN$ vertex and consider various prescriptions for restoring
current conservation.

As a second step, in
order to obtain the nuclear coincidence cross section or hadronic responses
one also needs the polarized spectral function, which depends on the target
polarization, to be multiplied by the polarized single-nucleon cross
section/responses. This is written in terms of a partial momentum
distribution, that depends on the particular initial and daugther nuclear
states. In this work we present results for the case of a typical medium
nucleus that can be polarized, namely the $^{39}$K nucleus, 
which is described
in shell model as a proton hole in the $d_{3/2}$ shell of $^{40}$Ca.

The general structure of the paper is as follows. In Section 2 we briefly
review the basic formalism for $\vec{A}(\vec{e},e'\vec{p})$ reactions focusing
on the PWIA. Here we evaluate the
single-nucleon tensor and analyze its dependence on the spin variables. In
Section 3 we present the results: first we focus on the single-nucleon
responses, where off-shell effects and symmetry properties are carefully
studied; second we present our predictions for 
final observables, namely  ``skewed'' polarization asymmetries, for $^{39}$K. 
Finally in Section 4 we summarize our conclusions.
The detailed expressions of the single nucleon tensors, responses and
momentum distribution are given in the Appendices.


\section{Formalism of $\vec{A}(\vec{e},e'\vec{p})B$ reactions}
\label{sec:formalism}


\begin{figure}
\begin{center}
\begin{picture}(300,200)(0,0)
\Text(80,177)[l]{$e$}
\ArrowLine(90,167)(140,133)
\Text(50,147)[l]{$K^{\mu}=(\varepsilon,\mathbf{k})$}
\Text(200,177)[l]{$e'$}
\ArrowLine(140,133)(190,167)
\Text(170,147)[l]{$K'^{\mu}=(\varepsilon',\mathbf{k'})$} 
\Vertex(140,133){1.5}
\Photon(140,133)(170,81){4}{4.5}
\Text(85,113)[l]{$Q^{\mu}=(\omega,\mathbf{q})$}
\ArrowLine(170,81)(220,48)
\ArrowLine(168,79)(218,46)
\Text(227,50)[l]{$B$}
\ArrowLine(120,48)(170,81)
\ArrowLine(118,50)(168,83)
\ArrowLine(122,46)(172,79)
\Text(110,38)[l]{A}
\Text(80,78)[l]{$P^{\mu}_A=(M_A,\mathbf{0})$}
\GCirc(170,81){6}{0}
\ArrowLine(170,81)(230,85)
\Text(235,80)[l]{$N$}
\Text(205,95)[l]{$P^{\mu}_{N}=(E_{N},\mathbf{p_{N}})$}
\Text(160,38)[l]{$P^{\mu}_{B}=(E_{B},\mathbf{p_{B}})$}
\end{picture}
\end{center}
\caption{\label{diagram}Feynman diagram for the $A(e,e'N)B$ process within 
the Born approximation.}
\end{figure}
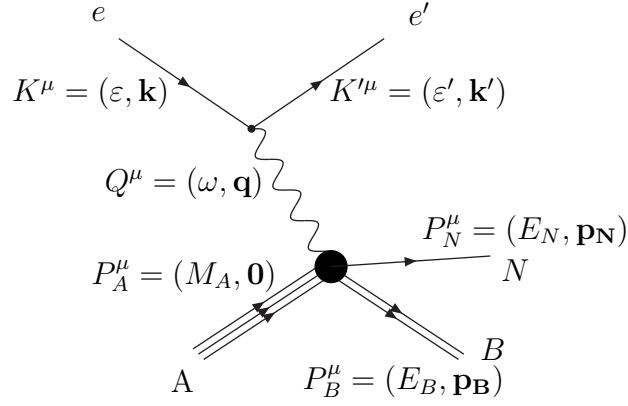

In this section we summarize the basic formalism involved in the
description of coincidence $(e,e'p)$ reactions~\cite{RaDo89}. The Feynman
diagram within the Born approximation (one virtual photon exchange) is
depicted in Fig.~\ref{diagram} and defines our conventions on energies and
momenta. In what follows we introduce the kinematical variables that
completely specify the process and use energy-momentum conservation to
inter-relate the various energies and momenta: \ba \nq &=& \nk-\nk' =
\np_N+\np_B \label{eq1s2}
\\
\omega &=& \varepsilon-\varepsilon' = E_B+E_N-M_A \, ,
\label{eq2s2}
\ea
where the target is assumed to be at rest in the laboratory frame.
The missing momentum $\np$ is defined as
\be
\np\equiv -\np_B=\np_N-\nq \, .
\label{eq3s2}
\ee

The general form for the coincidence cross section in the laboratory system is
\begin{equation}
\frac{d\sigma }{d\Omega _{e} d\varepsilon' d\Omega _{N}}=
\frac{2\alpha ^{2}}{Q^{4}}\left(\frac{\varepsilon'}{\varepsilon}\right)K
f_{rec}^{-1}\eta_{\mu\nu}W^{\mu\nu} \, ,
\label{eq1s3}
\end{equation}
where $K=p_NM_NM_B/M_A$, $\alpha$ is
the fine structure constant, $f_{rec}$ is the usual recoil
factor~\cite{RaDo89}, $\eta_{\mu\nu}$ is the leptonic tensor that can be
evaluated using trace techniques~\cite{RaDo89}, and $W^{\mu\nu}$ is
the hadronic tensor containing all of the nuclear structure and dynamics
information.  The latter is given in terms of the nuclear
electromagnetic transition currents in momentum space. Note that the above
equation is completely general and may contain all polarization degrees of
freedom. The electron beam polarization occurs in the antisymmetric part of
the leptonic tensor $\eta_{\mu\nu}$,
 while the target and recoil nucleon
polarizations enter through the hadronic tensor $W^{\mu\nu}$.

Using the general properties of the leptonic tensor, the contraction of the
leptonic and hadronic tensors can be decomposed in terms of leptonic
kinematical ``super-Rosenbluth'' factors and response functions. The
differential cross section can then be written as 
\begin{equation}
\frac{d\sigma}{d\Omega_e d\varepsilon' d\Omega_N}
=\Sigma+ h \Delta \,,
\label{responses}
\end{equation}
being
\begin{eqnarray}
\Sigma
&=&
 K\sigma_{M}f_{rec}^{-1}
\left(v_LR^L+v_TR^T+v_{TL}R^{TL}+v_{TT}R^{TT}\right)
\nonumber\\
\Delta
&=&
 K\sigma_{M}f_{rec}^{-1}
\left(v_{T'}R^{T'}+v_{TL'}R^{TL'}\right)\,,
\end{eqnarray}
 where $\sigma_M$ is the Mott cross section. The kinematic factors
$v_\alpha$ contain all of the dependence on the leptonic vertex aside from
overall multiplicative factors (see~\cite{RaDo89} for their explicit
expressions in the extreme relativistic limit). The factor $h=\pm 1$ is
the incident electron's helicity. The hadronic current enters via the response
functions $R^\alpha$, which thus contain all the information on the target and
ejected nucleon spin dependence.  The labels $L$ and $T$ refer to projections
of the current matrix elements longitudinal and transverse to the virtual
photon direction, respectively.


\subsection{Plane-Wave Impulse Approximation (PWIA)}


The PWIA constitutes the simplest approach to describing the reaction
mechanism for coincidence electron scattering reactions. It has been discussed
in detail in previous work~\cite{Frul85,Kel96,Bof96,Cab93}, and thus here we
simply summarize the basic expressions needed for the discussion to follow.
The basic assumptions in PWIA are the following: i) the electromagnetic
current is taken to be a one-body operator (impulse approximation), ii) the
ejected nucleon is a plane wave, i.e., the nucleon emerges from the nucleus
without interaction with the residual nuclear system, and iii) the nucleon
detected in the coincidence reaction is the one to which the virtual photon is
attached.  Thus, the hadronic final state in PWIA is simply characterized by
the product of the state of the residual nucleus $|B\rangle$ and the
on-shell knocked-out nucleon spinor state $|\np_N,S_N\rangle$. Moreover, the
momentum of the struck nucleon
coincides with the missing momentum defined in Eq.~(\ref{eq3s2}).
The hadronic tensor can then be written in the general form:
\be
W^{\mu\nu}=
\sum_{mm'}{\cal W}^{\mu\nu}_{mm'}(\np,\nq;S_N)M_{mm'}(\np,\nOmega^*) \, ,
\label{PWIA}
\ee 
where $S_N$ refers to the ejected nucleon four-spin and $\nOmega^*$ defines
the target polarization direction.
The tensor in (\ref{PWIA}) factorizes into a part that depends directly on the
$\gamma NN$ vertex and a part containing the nuclear structure dependence. The
former is the single-nucleon tensor given by
\be
{\cal W}^{\mu\nu}_{mm'}(\np,\nq;S_N)=
\langle\np_N,m_N(S_N)|\hat{\Gamma}^\mu|\np,m'(\xi)\rangle^\ast
\langle\np_N,m_N(S_N)|\hat{\Gamma}^\nu|\np,m(\xi)\rangle \, ,
\label{eq12s3}
\ee 
where $|\np,m(\xi)\rangle$ represents an on-shell spinor state
characterized by four-momentum 
$\overline{P}^\mu=\left(\overline{E}\equiv\sqrt{\np^2+m_N^2},\np\right)$ 
and spin four-vector
$\xi^\mu$, 
and $\hat{\Gamma}^\mu$ is the off-shell $\gamma NN$ vertex
operator.  The indices $m,m'$ denote the spin projection ($\pm 1/2$) in the
$z$ direction of the bound nucleon in its rest frame. The latter term 
\be 
M_{mm'}(\np,\nOmega^*)=\sum_{M_B}\langle
J_B M_B|a_{\np m'}|J_A J_A(\nOmega^*)\rangle^\ast 
\langle J_B M_B|a_{\np m}|J_A J_A(\nOmega^*)\rangle \, 
\label{spectral}
\ee 
is the polarized density matrix in momentum space for the polarized
target, assuming 100\% polarization in the $\nOmega^*$ 
direction and a specific nuclear transition $J_A\to J_B$ 
(see refs.~\cite{Cab93,Cab94,Edu95,Ama02} for details).

Introducing these results in the general expression (\ref{eq1s3}) 
for the cross section we obtain
\begin{equation}
\frac{d\sigma}{d\Omega_e d\varepsilon' d\Omega_N}=
Kf_{rec}^{-1}\sum_{mm'}\sigma^{ep}_{mm'}(\np,\nq;S_N)
M_{mm'}(\np,\nOmega^*) \, , \label{cross}
\end{equation}
where we have introduced the off-shell polarized electron-proton cross
section, $\sigma^{ep}_{mm'}$, that can be decomposed into polarized
single-nucleon response functions according to 
\ba
\sigma^{ep}_{mm'}(\np,\nq;S_N)
&=&
\frac{2\alpha^2}{Q^4}\left(\frac{\varepsilon'}{\varepsilon}\right)
\eta_{\mu\nu} {\cal W}^{\mu\nu}_{mm'}(\np,\nq;S_N)
\nonumber\\
&=& 
\sigma_M\left[\sum_{\alpha}v_\alpha{\cal R}_{mm'}^\alpha(\np,\nq;S_N) +h
  \sum_{\alpha'}v_{\alpha'}{\cal R}_{mm'}^{\alpha'}(\np,\nq;S_N)\right] \,
\label{sncspwia}
\ea with $\alpha=L,T,TL,TT$ referring to the electron-unpolarized responses
and $\alpha'=T',TL'$ to the electron-polarized ones. The explicit expressions
of the polarized single-nucleon responses as components of the single-nucleon
tensor ${\cal W}^{\mu\nu}_{mm'}$ are given in~\cite{Cab93}. It is also
important to point out that the functions ${\cal R}^{\alpha(\alpha')}_{mm'}$
depend in general on both nucleonic polarization 
degrees of freedom, namely the spin projections of the struck nucleon on the laboratory
$z$-axis, $m,m'$, and the ejected proton $S_N^\mu$. 
The hadronic response
functions in PWIA are then given simply by taking the combination \be R^
{\alpha(\alpha')}=\sum_{mm'}{\cal R}^{\alpha(\alpha')}_{mm'} (\np,\nq;S_N)
M_{mm'}(\np,\nOmega^*)\, . \label{hadronic} \ee

The polarized momentum 
distribution $M$ can be analyzed by introducing its scalar $\overline{M}$ and
vector $\widehat{\nM}$ spin components, which are defined as
follows~\cite{Cab94,Edu95} 
\ba
\overline{M}(\np,\nOmega^*) 
&=& 
\textrm{Tr}\left[M(\np,\nOmega^*)\right] \label{scalar} \\
\widehat{\nM}(\np,\nOmega^*)
& =& 
\textrm{Tr}\left[\nsigma
 M(\np,\nOmega^*)\right] \label{vector} \, .  
\ea

In this paper we first focus our attention on the single-nucleon part of the
problem, generalizing our previous results for $A(\vec{e},e'\vec{p})$ and
$\vec{A}(\vec{e},e'p)$ processes. Before entering into a detailed discussion
of the single-nucleon aspect of the problem, it is important to point out that
the target polarization dependence occurs in the vector as well as in the
scalar components of the momentum distributions. The analysis of this
dependence is further simplified by writing the density matrix in terms of
different tensor polarization components: $M_{mm'}=\sum_I M_{mm'}^{(I)}$.  As
shown in~\cite{Cab94} the scalar term $\overline{M}$ only gets contributions
from even-rank tensors, while only odd-rank tensors contribute to
$\widehat{\nM}$.  The totally unpolarized target situation corresponds to
taking $I=0$, thus, only the scalar density $\overline{M}$ occurs in this
case.

\subsection{The Single-Nucleon Tensor: General Remarks}

Following closely the notation introduced in~\cite{Cab93} in the case of
$\vec{A}(\vec{e},e'p)$ reactions, here we develop the explicit expression for
the appropriate single-nucleon tensor that enters in the analysis of
$\vec{A}(\vec{e},e'\vec{p})$. Given a specific prescription for the off-shell
vertex $\hat{\Gamma}^\mu$ (we will consider the standard ones introduced by de
Forest~\cite{For83}) and making use of trace techniques, the single-nucleon
tensor reads 
\be {\cal W}^{\mu\nu}_{mm'}(\np,\nq;S_N)
=\frac{1}{16M_N^2}
\textrm{Tr}
\left[
  (\overline{\pslash}+M_N)(\delta_{mm'}+ \gamma_5\cslash_{mm'})
  \gamma^0\hat{\Gamma}^{\mu\dag}\gamma^0(\pnslash+M_N)(1+\gamma_5\snslash)
  \hat{\Gamma}^\nu
\right]
\label{sntensor}
\ee 
with the pseudovector matrix
$\xi_{mm'}^\mu\equiv\langle\np,m'(\xi)|\gamma^\mu\gamma^5|\np,m(\xi)\rangle$
which reduces to the four-spin $\xi^\mu$ in the diagonal case $m=m'$.  
  The tensor (\ref{sntensor}) can be split into a symmetric (${\cal S}$) and
  an antisymmetric (${\cal A}$) term under $\mu \leftrightarrow \nu$,
  according to the number (even or odd) 
  of $\gamma_5$ matrices appearing in the trace. One
  can write in general 
\be 
{\cal W}^{\mu\nu}_{mm'}(\np,\nq;S_N)
=
\delta_{mm'}\left[{\cal S}^{\mu\nu}
+ i{\cal  A}^{\mu\nu}(S_N)\right]+{\cal S'}^{\mu\nu}_{mm'}(S_N)
+ i{\cal A'}^{\mu\nu}_{mm'} \, , 
\label{sntensor1} 
\ee 
where the dependence upon the outgoing nucleon spin has been explicitly
indicated.  Moreover, while ${\cal S}^{\mu\nu}$ and ${\cal A}^{\mu\nu}(S_N)$
are real, the tensors ${\cal A'}^{\mu\nu}_{mm'}$ and ${\cal
  S'}^{\mu\nu}_{mm'}(S_N)$ are real for diagonal components $m=m'$ and in
general complex for off-diagonal terms.

The expression (\ref{sntensor1}) contains the whole information on the
polarization degrees of freedom of the problem and shows clearly the
difference with the cases studied in refs.~\cite{Cab93,Cris02}.  So, for a
scattering process off polarized targets, if the outgoing nucleon polarization
is not measured, the only tensors involved are ${\cal S}^{\mu\nu}$ and ${\cal
  A'}^{\mu\nu}_{mm'}$: this corresponds to the analysis presented in
\cite{Cab93}.  On the contrary, for a process with an unpolarized target and
polarized ejected nucleon, the single-nucleon tensor reduces to ${\cal
  S}^{\mu\nu}+i{\cal A}^{\mu\nu}(S_N)$, i.e., the whole dependence on the
recoil nucleon spin is contained in the antisymmetric tensor, which means that
only the electron-polarized hadronic responses $R^{T'}$ and $R^{TL'}$ depend
upon the nucleon polarization.  As known, this is a consequence of the PWIA
assumption. In the totally unpolarized situation, only ${\cal S}^{\mu\nu}$
survives giving rise to the four electron-unpolarized responses $R^\alpha$,
$\alpha=L,T,TL,TT$. Finally, it is important to note that the symmetric term
${\cal S'}^{\mu\nu}_{mm'}(S_N)$ depends simultaneously on both the target and
ejected nucleon polarizations and contributes to those responses which are
only constructed from the symmetric tensor, i.e., the four
electron-unpolarized ones. In this paper our
main attention will be focused on these new contributions giving rise to
``skewed'' polarization, as shown below.

In order to evaluate the various components of 
${\cal   W}^{\mu\nu}_{mm'}(\np,\nq;S_N)$ 
we make use of the spin precession technique
developed in~\cite{Cab93}. This method essentially amounts to first calculate
the tensor
\ba
{\cal W}^{\mu\nu}(\xi;S_N)
&\equiv&
{\cal W}^{\mu\nu}(\theta_R,\phi_R;S_N)
\nonumber \\
&=&
\frac{1}{16M_N^2}\textrm{Tr}\left[
  (\overline{\pslash}+M_N)(1+\gamma_5\xislash)\gamma^0\Gamma^{\mu\dag}\gamma^0
  (\pnslash+M_N)(1+\gamma_5\snslash)\Gamma^\nu\right] 
\nonumber \\
&\equiv& 
\underbrace{ 
  {\cal S}^{\mu\nu}
 +{\cal S'}^{\mu\nu}(\theta_R,\phi_R;S_N)}_{{\cal S}^{\mu\nu}(\xi;S_N)}+
i\underbrace{
  \left[{\cal A}^{\mu\nu}(S_N)
      +{\cal A'}^{\mu\nu}(\theta_R,\phi_R)
  \right]}_{{\cal A}^{\mu\nu}(\xi;S_N)}
  \, ,
\label{sntensor2} 
\ea 
where the angles $(\theta_R,\phi_R)$ specify the spin direction of the hit nucleon
(with respect to the quantization axis) in its rest frame.
Obviously, when boosted to the laboratory
frame, the four-spin $\xi^\mu$ precesses to the positive $(\theta_L,\phi_L)$ direction 
(see~\cite{Cab93} for details on how the connection is made). 
Following \cite{Cab93} and after
evaluating the components of $\xi^\mu$, one finds the relations: 
\ba 
{\cal  W}^{\mu\nu}_{++}
&=&
{\cal S}^{\mu\nu}+{\cal S'}^{\mu\nu}(0,0;S_N)+
i\left[{\cal A}^{\mu\nu}(S_N)+{\cal A'}^{\mu\nu}(0,0)\right] \nonumber 
\\
{\cal W}^{\mu\nu}_{--}
&=&
{\cal S}^{\mu\nu}-{\cal S'}^{\mu\nu}(0,0;S_N)+
i\left[{\cal A}^{\mu\nu}(S_N)-{\cal A'}^{\mu\nu}(0,0)\right] 
\nonumber \\
{\cal W}^{\mu\nu}_{+-}
&=&
{\cal S'}^{\mu\nu}(\frac{\pi}{2},0;S_N) -{\cal
  A'}^{\mu\nu}(\frac{\pi}{2},\frac{\pi}{2})+ i\left[{\cal
    A'}^{\mu\nu}(\frac{\pi}{2},0)+
  {\cal S'}^{\mu\nu}(\frac{\pi}{2},\frac{\pi}{2};S_N)\right] 
\nonumber \\
{\cal W}^{\mu\nu}_{-+}
&=&
{\cal S'}^{\mu\nu}(\frac{\pi}{2},0;S_N) +{\cal
  A'}^{\mu\nu}(\frac{\pi}{2},\frac{\pi}{2})+ i\left[{\cal
    A'}^{\mu\nu}(\frac{\pi}{2},0)- {\cal
    S'}^{\mu\nu}(\frac{\pi}{2},\frac{\pi}{2};S_N)\right] \, ,
\label{components}
\ea
where we have used the results ${\cal S'}^{\mu\nu}(\pi,0;S_N)=
-{\cal S'}^{\mu\nu}(0,0;S_N)$ and 
${\cal A'}^{\mu\nu}(\pi,0)=-{\cal A'}^{\mu\nu}(0,0)$.

The analysis further simplifies by defining a new basis in which the
single-nucleon tensors are given in the form~\cite{Cab94,Edu95}:
\ba
&& 0:\frac{1}{2}\left[{\cal W}^{\mu\nu}_{++}+{\cal W}^{\mu\nu}_{--}
\right]={\cal S}^{\mu\nu}+i{\cal A}^{\mu\nu}(S_N) \nonumber \\
&& z:\frac{1}{2}\left[{\cal W}^{\mu\nu}_{++}-{\cal W}^{\mu\nu}_{--}
\right]={\cal S'}^{\mu\nu}(0,0;S_N)+i{\cal A'}^{\mu\nu}(0,0) \nonumber \\
&& x:\frac{1}{2}\left[{\cal W}^{\mu\nu}_{+-}+{\cal W}^{\mu\nu}_{-+}
\right]={\cal S'}^{\mu\nu}(\frac{\pi}{2},0,S_N)
         +i{\cal A'}^{\mu\nu}(\frac{\pi}{2},0) \nonumber \\
&& y:-\frac{i}{2}\left[{\cal W}^{\mu\nu}_{+-}-{\cal W}^{\mu\nu}_{-+}
\right]= {\cal S'}^{\mu\nu}(\frac{\pi}{2},\frac{\pi}{2};S_N)+i
         {\cal A'}^{\mu\nu}(\frac{\pi}{2},\frac{\pi}{2})
          \, .
\label{basis}
\ea
Note that $z$ 
denotes the component along the transfer momentum $\nq$, whereas
$x$ and $y$ are the vector components in the perpendicular scattering plane. 

The explicit calculation of the single-nucleon tensor ${\cal
  W}^{\mu\nu}(\xi;S_N)$ for the two common choices, CC1 and CC2, of the
current operator is presented in Appendix A. Once the single-nucleon tensor is
known, the various responses 
${\cal  R}^{\alpha(\alpha')}_{mm'}(\theta_R,\phi_R;S_N)$ 
are constructed directly by
taking the appropriate components.  In Appendix B we summarize the expressions
of all the single-nucleon responses making use of the basis introduced
in~(\ref{basis}). Hence, by analogy with the polarized density and
following~\cite{Edu95}, the single-nucleon spin dependent responses ${\cal
  R}^{\alpha(\alpha')}_{mm'}$ can be analyzed by introducing scalar
$\overline{{\cal R}}^{\alpha}$, $\overline{{\cal R}}^{\alpha'}(S_N)$ and
vector $\widehat{\ncalR}^{\alpha}(S_N)$, $\widehat{\ncalR}^{\alpha'}$
components.  As usual, $\alpha=L,T,TL,TT$ and $\alpha'=T',TL'$. Note that the
vector responses (given by the three components referred to the $xyz$ frame)
only enter when the target is polarized, whereas the scalar ones may
contribute even if the target is unpolarized (zero rank tensor in the scalar
momentum distribution). Moreover, the explicit dependence in the outgoing
nucleon polarization only enters in the scalar responses $\overline{\cal
  R}^{\alpha'}(S_N)$, which give rise to the two polarized hadronic responses
$R^{T'}$ and $R^{TL'}$ that contribute to $A(\vec{e},e'\vec{p})$ reactions,
and in the vector terms $\widehat{\ncalR}^{\alpha}(S_N)$, which also require
polarization of the target and only affect the electron-unpolarized responses
$R^L,R^T,R^{TL}$ and $R^{TT}$.  

\subsection{Skewing responses and asymmetries}

The hadronic response functions (\ref{hadronic}) can be expressed as follows:
\begin{itemize}
\item Electron-unpolarized responses: $R^\alpha$ with $\alpha=L,T,TL,TT$
\be
R^\alpha=
\overline{\cal R}^\alpha \overline{M}(\np,\nOmega^*)+
\widehat{\ncalR}^\alpha(S_N)\cdot\widehat{\nM}(\np,\nOmega^*) \, .
\label{resp1}
\ee
\item Electron-polarized responses: $R^{\alpha'}$ with $\alpha'=T',TL'$
\be
R^{\alpha'}=
\overline{\cal R}^{\alpha'}(S_N) \overline{M}(\np,\nOmega^*)+
\widehat{\ncalR}^{\alpha'}\cdot\widehat{\nM}(\np,\nOmega^*) \, .
\label{resp2}
\ee
\end{itemize}
From~(\ref{resp1}) and~(\ref{resp2}) it comes out
that for reactions with unpolarized targets, only the two terms involving the
scalar momentum distribution (for zero rank tensor) contribute. From these,
the one in~(\ref{resp1}) refers to the fully unpolarized situation whereas
that in~(\ref{resp2}) requires measuring the polarization of the ejected
nucleon as well.  In the case of polarized targets, the vector momentum
distribution also enters. Note, however, that in the case of the four
electron-unpolarized responses~(\ref{resp1}) such contribution only enters if
the outgoing nucleon spin is also measured. 

The expressions given by
eqs.~(\ref{resp1},\ref{resp2}) constitute one of the basic outcomes of this
work because they show in a very clear way how the polarization degrees of
freedom get organized in the scattering process.  Therefore, concerning the
single-nucleon aspect of the problem, a crucial difference emerges in the
electron-polarized and electron-unpolarized observables. In the former, the
whole spin dependence can be decoupled into two separate single nucleon
responses, each one depending exclusively on one of the nucleon spin
variables, the outgoing nucleon or the target (bound nucleon).  This result is in
accordance with the form in which the four-spin variables enter in the
antisymmetric tensor, i.e., $(\xi\pm S_N)^\mu$ (see Appendix A).  The
electron-unpolarized case clearly differs: here the whole spin dependence in
$\widehat{\ncalR}^\alpha(S_N)$ comes from the symmetric tensor, where the
four-spin vectors enter through their scalar product (i.e., $S_N\cdot\xi$),
 and involves the polarization of both the target and the emitted nucleon.
Therefore, since results for the totally unpolarized and
  polarized recoil nucleon responses, $\overline{\cal R}^\alpha$ and
  $\overline{\cal R}^{\alpha'}(S_N)$, have been presented in~\cite{Cris02},
  whereas the pure polarized target single-nucleon responses
  $\widehat{\ncalR}^{\alpha'}$ were studied in~\cite{Cab93}, the analysis in
  this work will be focused on the new responses
  $\widehat{\ncalR}^\alpha(S_N)$.
  Accordingly, in the
following we restrict to the case of unpolarized electrons, for which
only the $\Sigma$ part of the general cross section (\ref{responses}) enters.

The differential cross section (\ref{cross})
for $\vec{A}(e,e'\vec{p})$ can be written in PWIA in the form
\be
\frac{d\sigma}{d\Omega_ed\varepsilon'd\Omega_N}= \Sigma =
Kf_{rec}^{-1}\left[\overline{\sigma}^{ep}\overline{M}(\np,\nOmega^*)+
\widehat{\nsigma}^{ep}(S_N)\cdot\widehat{\nM}(\np,\nOmega^*)\right] \, ,
\label{diff_cross}
\ee
where the scalar and vector single-nucleon cross sections 
\begin{eqnarray}
\overline{\sigma}^{ep}
&=&
\sigma_M\sum_{\alpha}v_\alpha \overline{\cal R}^\alpha 
\label{snscalar} \\
\widehat{\nsigma}^{ep}(S_N)
&=&
\sigma_M\sum_{\alpha}v_\alpha \widehat{\ncalR}^\alpha(S_N) 
\label{snvector}
\end{eqnarray}
have been introduced in analogy with the single nucleon response functions.

In the next section we present predictions for the polarization asymmetries
obtained from the difference of cross sections for
opposite ejected nucleon spin directions (denoted by the three-vector $\ns_N$) 
divided by their sum, for a selected target orientation:
\begin{equation}\label{asimetria-exp}
A_{\ns_N}=\frac{\Sigma(\ns_N)-\Sigma(-\ns_N)}{\Sigma(\ns_N)+\Sigma(-\ns_N)}\,.
\end{equation}
Making use of (\ref{diff_cross}) and taking into account that the whole
dependence of $\widehat{\nsigma}^{ep}$ upon the ejected nucleon spin components 
is linear (see Appendix A), the polarization asymmetries can be computed as 
\begin{equation}
A_{\ns_N}=\frac{\widehat{\nsigma}^{ep}(\ns_N)\cdot
  \widehat{\nM}(\np,\nOmega^*)}
{\overline{\sigma}^{ep}\overline{M}(\np,\nOmega^*)} \, .
\label{asymmetry} 
\end{equation} 

Finally, to make connection with the concept of ``skewed'' recoil polarization,
we provide a geometrical interpretation of the polarized momentum distribution by 
introducing the following vector field
\begin{equation}\label{spin}
\nchi(\np,\nOmega)\equiv
\frac{\widehat{\nM}(\np,\nOmega^*)}{\overline{M}(\np,\nOmega^*)}.
\end{equation}
It can be shown that in the independent particle model, this is a
unit vector, $|\nchi|=1$, and that, semi-classically, it coincides with
the local spin field in momentum space~\cite{Ama99,Ama02,Ama04}
(see an example in Appendix C for the particular case of the 
$^{39}$K nucleus analyzed in the next section).
Since the dependence of the cross section on the final spin direction,
$S^\mu_N$, is linear, one can write the $i$-th component ($i=x,y,z$) of the vector
cross section in the general form
\begin{equation}
\widehat{\sigma}_i^{ep}(\ns_N)= 
\widehat{\sigma}_{\mu i}^{ep}S_N^\mu \, ,
\end{equation}
where we have introduced the tensor components 
$\widehat{\sigma}_{\mu i}^{ep}$ of the ``skewing'' cross section. 
Note that the half-covariant form 
of this tensor, with a covariant index, $\mu$, and a spatial one, $i$, 
reflects the fact that the target momentum distribution is 
being described within a non-relativistic framework. 
Hence the full covariance of the
equations is broken at this level, since we do not provide a
covariant description of the nuclear target.

Similarly to the vector cross-section, the asymmetry (\ref{asymmetry}) can also be written as
\begin{equation}
A_{\ns_N}=A_{\mu i}S_N^\mu \chi_i \, ,
\end{equation}
where a sum over the spatial index, $i$, is understood,
and we have introduced the tensor asymmetry
\begin{equation}
A_{\mu i} = 
\frac{ \widehat{\sigma}_{\mu i}^{ep}}{ \overline{\sigma}^{ep}} \, .
\end{equation}
Now for given kinematics and target polarization, we introduce the ``skewed'' four-vector
$R^\mu=(r_0,\nr)$, defined as
\begin{equation} \label{wrenched-vector}
R_\mu = -A_{\mu i} \chi_i \, .
\end{equation}
The asymmetry can then be written as
\begin{equation}
A_{\ns_N} = - R_{\mu}S_N^\mu
= -r_0\sqrt{\ns_N^2-1}+\nr\cdot\ns_N
\end{equation}
since $S_N$ verifies $S_N^2=-1$. 
For a fixed value of $R^\mu$,
the above asymmetry is maximum for $\ns_N$ in the direction
of $\nr$ and $\ns_N^2 = \nr^2/(\nr^2-r_0^2)$.
Therefore $R^\mu$ determines the preferred
polarization direction for the final proton: it is 
obtained as the linear transformation (\ref{wrenched-vector})
of the local spin field (\ref{spin}), producing in this way the
``skewing'' effect we are referring to in this work.


\section{Results for $\vec{A}(e,e'\vec{p})$}


In this section we present results for the new ``skewing'' observables 
arising in $\vec{A}(e,e'\vec{p})$ reactions. Two
kinematical settings already used in previous studies~\cite{Cab93,Cris02} are
considered. The former corresponds to quasi-perpendicular kinematics,
where the values of the transfer momentum $q$ and transfer energy $\omega$ are
fixed. The values selected are $q=500$ MeV/c and $\omega=131.6$ MeV, which
corresponds almost to the quasielastic peak center. The azimuthal angle of the
recoil nucleon momentum is fixed to $\phi_N=0^0$, i.e., coplanar kinematics.
The second kinematics corresponds to parallel kinematics, where the outgoing
nucleon momentum $\np_N$ is parallel to $\nq$, i.e., $\theta_N=0^o$. The
kinetic energy of the recoil nucleon is fixed, as well as the electron beam
energy.

Concerning the recoil nucleon polarization, we 
use the coordinate system defined by the axes $\nl$ (parallel to
$\np_N$), $\nn$ (perpendicular to the plane containing $\np_N$ and $\nq$) and
$\ns$ (determined by $\nn\times\nl$). This contrasts with the target
polarization, which is referred to the $(x,y,z)$
system with $z$ given by the direction of $\nq$. Hence, an important
difference emerges when analyzing results corresponding to both kinematics.
For the quasi-perpendicular case, varying the missing momentum means
changing the direction in which the outgoing nucleon is detected. Hence the
$\nl$ and $\ns$ directions which specify the recoil nucleon spin do not
coincide with the components $z,x$ used for the bound nucleon polarization
components. On the contrary, the transverse direction $\nn$ points along the
$y$ axis as far as coplanar, $\phi_N=0^o$, kinematics is selected. Within the
parallel situation, varying the missing momentum $p$ implies changing the
magnitude of $q$.  In this case the system specified by
$(\nl,\ns,\nn)$ coincides with $(z,x,y)$. As will be shown, this leads to a
cancellation of some responses and, moreover, symmetries in most of the
scalar and vector single-nucleon responses emerge.

\begin{figure}[tbp]
\begin{center}
\includegraphics[scale=0.8, bb= 100 280 500 740]{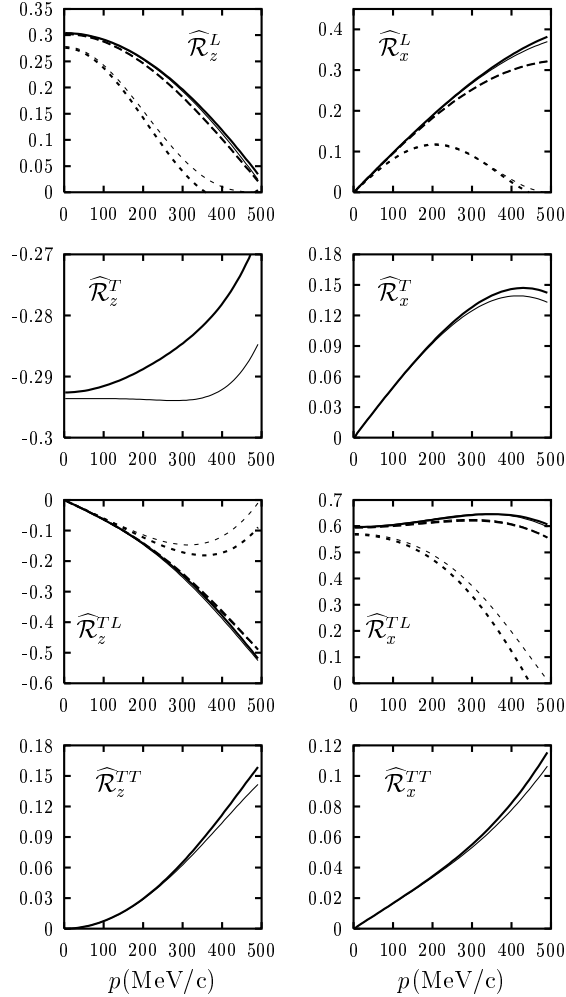}
\end{center}
\caption{Electron-unpolarized vector single-nucleon responses
$\widehat{\ncalR}^\alpha(S_N)=(\widehat{\cal R}^\alpha_x,\widehat{\cal
R}^\alpha_y, \widehat{\cal R}^\alpha_z)$ for quasi-perpendicular
kinematics at momentum and energy transfer $q=500$ MeV/c and
$\omega=131.6$ MeV, respectively.  The outgoing nucleon polarization
is along the longitudinal direction ($\nl$).  
Thin lines correspond to the CC1 current operator and thick lines to the
CC2 current. Results are shown for Landau (solid lines), Coulomb 
(dashed lines) and Weyl (short-dashed lines) gauges.
\label{snresponses_perp_l}
}
\end{figure}

\subsection{Single-nucleon responses \label{sect}}

We start by showing the new vector polarized single-nucleon responses
$\widehat{\ncalR}^{\alpha}(S_N)$ that enter in the analysis of
$\vec{A}(e,e'\vec{p})$ reactions within PWIA. 
Following previous studies on
$\vec{A}(\vec{e},e'p)$ and $A(\vec{e},e'\vec{p})$
processes~\cite{Cab93,Cris02}, here the off-shell character of the
bound nucleon is explored making use of the current operator choice,
CC1 vs CC2, and several prescriptions to restore
 current conservation, namely Coulomb, Landau and
Weyl gauges (see~\cite{Cab93,Cab98a} for more details).  
The discrepancy among the various models of electromagnetic 
current allows one to get
some insight into the importance of the off-shell effects on these observables.

\begin{figure}
\begin{center}
\includegraphics[scale=0.8, bb= 100 280 500 740]{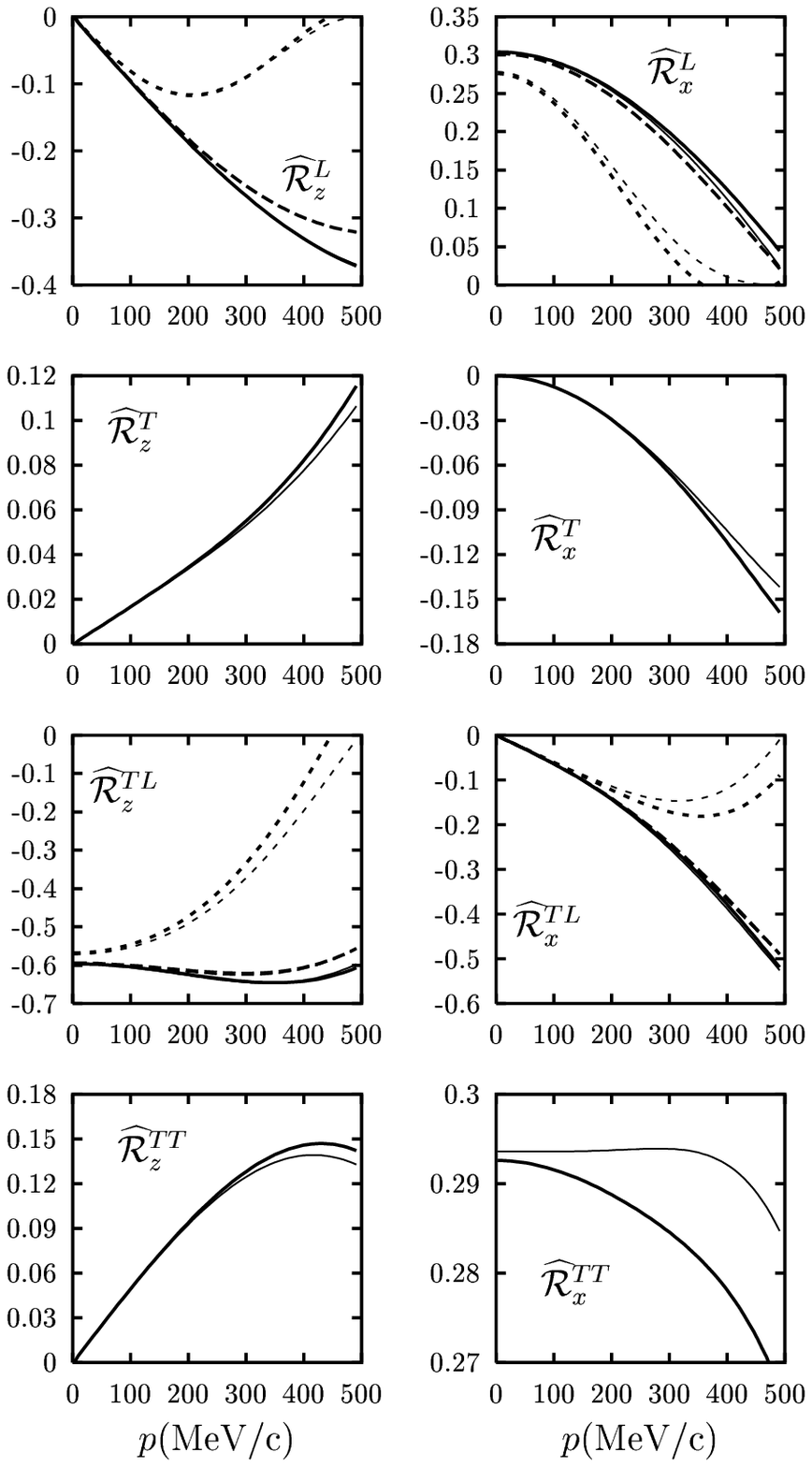}
\end{center}
\caption{Same as fig.~2 but for the outgoing nucleon polarized along
the sideways direction ($\ns$). \label{snresponses_perp_s} }
\end{figure}

\begin{figure}
\begin{center}
\rotatebox[origin = lb]{90}{\scalebox{3}[1]}
\includegraphics[bb= 100 510 500 740]{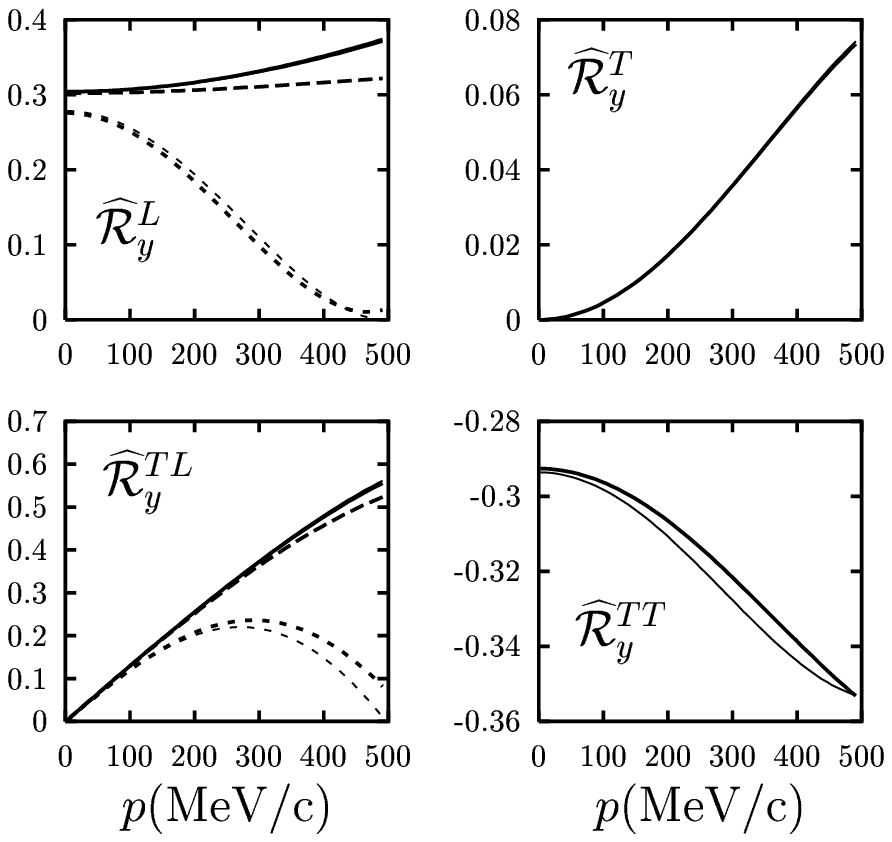}
\end{center}
\caption{Same as fig.~2 but for the outgoing nucleon polarized along the normal
direction ($\nn$).\label{snresponses_perp_n} 
}
\end{figure}

Figures~\ref{snresponses_perp_l}--\ref{snresponses_perp_n} show the components
of the vector $\widehat{\ncalR}^\alpha (S_N)$ for quasi-perpendicular
kinematics as functions of the missing momentum.  Specifically, results
in Fig.~\ref{snresponses_perp_l} correspond to longitudinal ($\nl$),
Fig.~\ref{snresponses_perp_s} to sideways ($\ns$), and
Fig.~\ref{snresponses_perp_n} to normal ($\nn$) recoil nucleon polarization.
In the first two cases, the components $\widehat{\cal R}^\alpha_y$ are zero
because of the azimuthal angle selected, $\phi_N=0^o$. On the contrary, for
outgoing polarization along $\nn$ (Fig.~\ref{snresponses_perp_n}), the only
components which survive are $\widehat{\cal R}^\alpha_y$.  Note that,
as already mentioned, the $\nn$ direction lies along the $y$ axis for $\phi_N=0^o$.

To begin the analysis we discuss the symmetries emerging from our
results, by comparing Figs.~\ref{snresponses_perp_l}
and~\ref{snresponses_perp_s}, i.e., recoil nucleon spin along $\nl$ and $\ns$,
respectively. These symmetries are  
\begin{eqnarray}
\widehat{\cal R}^{L(TL)}_z(\nl)=\widehat{\cal R}^{L(TL)}_x(\ns) &;& 
\widehat{\cal R}^{L(TL)}_x(\nl)=-\widehat{\cal R}^{L(TL)}_z(\ns)
\label{sym1}
\\
\widehat{\cal R}^{T(TT)}_z(\nl)=-\widehat{\cal R}^{TT(T)}_x(\ns)&;&
\widehat{\cal R}^{T(TT)}_x(\nl)=\widehat{\cal R}^{TT(T)}_z(\ns)\,.
\label{sym4}
\end{eqnarray}
Note the interchange between the two purely transverse responses.
An analytical proof of the above symmetries can be provided
within the non relativistic limit~\cite{notes}.

In the case of recoil nucleon spin along 
the normal direction, $\nn$, the polarized single-nucleon responses shown in
Fig.~\ref{snresponses_perp_n} result to be related to the scalar 
responses $\overline{\cal R}^\alpha$, corresponding to unpolarized recoil protons 
(see results in refs.~\cite{Cab93,Cris02}).  In
fact the following identities hold for all the off-shell prescriptions: 
\begin{eqnarray}
\widehat{\cal R}^{L(TL)}_y(\nn)=\overline{\cal R}^{L(TL)} &;&
\widehat{\cal R}^{T(TT)}_y(\nn)=-\overline{\cal R}^{TT(T)}\,.
\label{sym6}
\end{eqnarray}
From these specific symmetries one could think that no new
information could be obtained by measuring the nucleon polarization
in these conditions. However note that the different role played by the $T$
and $TT$ responses would make this measurement an alternative  way 
to separate the contributions of the different responses.
As an example let us consider the case of a $^{39}$K nucleus 
polarized in the $y$ direction and described in a simple shell model 
(see section 3.2 and appendix C). 
 In this situation and for coplanar kinematics,
only the $y$ component of the vector momentum distribution
survives. The single-nucleon contribution to the cross
section (\ref{diff_cross}) for $\ns_N$ along the $\nn$ direction is then given by
\begin{equation}
\overline{\sigma}^{ep}-\widehat{\sigma}_y^{ep}(\nn)
= (v_T+v_{TT})(\overline{\cal R}^T+ \overline{\cal R}^{TT}) \,,
\end{equation} 
since the $L$ and $TL$ responses cancel out.
This would allow to isolate the sum of the $T$ and $TT$ responses.
In general, FSI are expected to break these symmetries at
some level, since it is known that an additional normal polarization is 
induced even for unpolarized nuclei.
However for intermediate values of the momentum transfer considered here
($q\simeq 500$ MeV/c)
the polarization induced by FSI is expected to be small ($\leq 0.2$). 
This is indicated both by experimental data~\cite{Woo98} and by calculations~\cite{Kaz04,Ama04}.

Regarding the off-shell effects and gauge ambiguities in 
$\widehat{\ncalR}^\alpha(S_N)$, the
discrepancies among the different prescriptions observed in
Figs.~\ref{snresponses_perp_l}-\ref{snresponses_perp_n} are similar to
the ones already presented and discussed at length for the scalar
$\overline{\cal R}^\alpha,\overline{\cal R}^{\alpha'}(S_N)$ and vector
$\widehat{\ncalR}^{\alpha'}$ components \cite{Cab93,Cris02}.
The basic findings may be summarized as follows.  For those responses
that involve the longitudinal component of the current, i.e., $L$ and
$TL$, the largest deviations are observed for the Weyl gauge (short-dashed lines). On the
contrary, the four remaining off-shell prescriptions based on both
current operators and the Coulomb and Landau gauges lead to similar
results in all the cases. Gauge ambiguities do not affect the pure
transverse responses $T$ and $TT$, and the two current operator
choices give rise to results which do not deviate significantly.  

\begin{figure}
\begin{center}
\rotatebox[origin = lb]{90}{\scalebox{3}[1]}
\includegraphics[scale=0.8,  bb= 40 400 560 740]{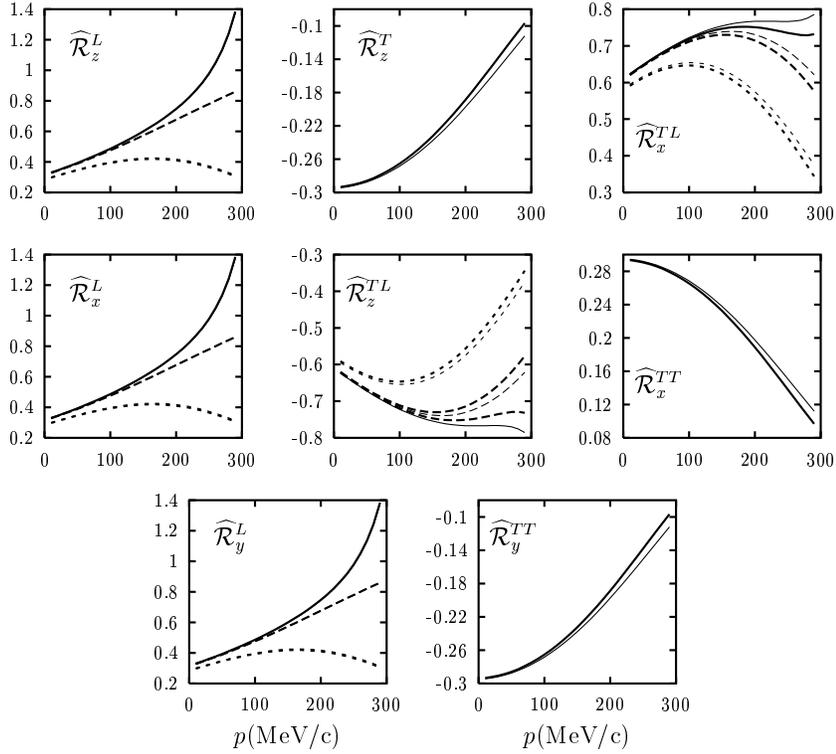}
\end{center}
\caption{Same responses as in Fig.~2 but for parallel kinematics. Top, middle
and bottom results correspond to the ejected nucleon polarized along the
$\nl$, $\ns$ and $\nn$ directions, respectively. The value of the outgoing
nucleon momentum is $p_N=490$ MeV/c and the electron beam energy 
$\varepsilon=500$ MeV.
\label{snresponses_paral} }
\end{figure}

Results in parallel kinematics are presented in Fig.~\ref{snresponses_paral}.
As known, in this situation the usual directions that specify the recoil
nucleon polarization $(\nl,\ns,\nn)$ coincide with the axes $(z,x,y)$ used to
characterize the bound nucleon (target) polarization. 
This leads to many symmetry relations among the various responses.
Moreover, many components in
the vector term $\widehat{\ncalR}^\alpha(S_N)$ vanish. Upper, middle and
bottom rows in Fig.~\ref{snresponses_paral} correspond to longitudinal,
sideways and normal ejected nucleon polarizations, respectively.  From the
total set of twelve $\widehat{\cal R}^\alpha_{x,y,z}$ components, only three
survive for $\nl$ and $\ns$ polarization directions, and two in the case of
normal $\nn$. Furthermore, the symmetries among different single-nucleon
responses, shown in eqs.~(\ref{sym1}--\ref{sym6}), 
also apply to the results in parallel
kinematics. Note that, in this particular situation, the only surviving
responses are those which differ from zero in the $p\rightarrow 0$ limit for
quasi-perpendicular kinematics. Obviously, the limit of missing momentum going
to zero in $(q,\omega)$-constant kinematics corresponds to
$\theta_N\rightarrow 0$, i.e., approaching parallel kinematics. Finally, the
two components for normal recoil nucleon polarization (bottom line) also
satisfy the relations: $\widehat{\cal R}^L_y(\nn)=\widehat{\cal
  R}^L_z(\nl)=\widehat{\cal R}^L_x(\ns)$ and $\widehat{\cal
  R}^{TT}_y(\nn)=\widehat{\cal R}^T_z(\nl)=-\widehat{\cal R}^{TT}_x(\ns)$.
Finally, although the two scalar responses that survive in parallel 
kinematics are not shown here, it is worth pointing out that they
verify the relations
$\overline{\cal R}^L=\widehat{\cal R}^L_y(\nn)$ and 
$\overline{\cal R}^T=-\widehat{\cal R}^{TT}_y(\nn)$. 
These symmetries are exactly fulfilled in all the off-shell 
prescriptions which, 
otherwise, lead to ambiguities of the same order as those discussed for 
quasi-perpendicular kinematics.

\subsection{Skewed polarization asymmetries}

In the previous section we have shown the ``skewing'' single nucleon
responses which are purely associated to the simultaneous measurement 
of target and ejected nucleon polarizations. In this section we 
explore polarized observables such as the cross sections and, 
more specifically, the polarization asymmetries introduced in~(\ref{asymmetry}).

The evaluation of the cross sections, within PWIA, requires not only the above
discussed single-nucleon ingredients but also the polarized momentum
distribution.  Being aware of the crucial role of FSI upon cross sections and
hadronic response functions, here we restrict our attention to polarization
asymmetries.  In fact ratios of two cross sections are expected to be
much less sensitive to FSI effects. This is the case for transferred
polarization asymmetries in $A(\vec{e},e'\vec{p})$ reactions, where the PWIA
results coincide for moderate missing momentum with much more elaborate
relativistic distorted wave impulse approximation (RDWIA)
calculations~\cite{Cris04}. The same comment applies to the analysis of
electron scattering on polarized targets~\cite{Edu95,Ama99,Ama02,Ama04}.

To analyze the polarization asymmetries  we consider again the
directions $\nl,\ns,\nn$ for  the ejected nucleon polarization,
and target orientation selected along the axes $x,y,z$.
A detailed study of the polarized momentum distribution for deformed nuclei
has been presented in~\cite{Cab94,Edu95}. The case of spherical nuclei within
a shell model framework has been analyzed in~\cite{Ama96} for inclusive
$\vec{A}(\vec{e},e')$ processes and in~\cite{Ama98b,Ama99,Ama02} 
for exclusive ones.
Here we consider a simple model where the last bound nucleons are responsible 
for the spin of the polarized target.  
Specifically for $^{39}$K, described as a hole nucleus with respect to the 
$^{40}$Ca core, the orbit involved in the process is $d_{3/2}$.  
Since the target has spin $\frac{3}{2}$, the scalar
polarized momentum distribution has even rank ($I=0,2$) components, while the
vector term $\widehat{\nM}$ odd rank ($I=1,3$) components. 
Only transitions
leading to the residual nucleus $^{38}$Ar in its ground state, described as two
protons in the $d_{3/2}$ orbit coupled to total angular momentum $J_B=0$, are
considered.

\begin{figure}
\begin{center}
\rotatebox[origin = lb]{90}{\scalebox{3}[1]}
\includegraphics[scale=0.8,  bb= 80 400 520 750]{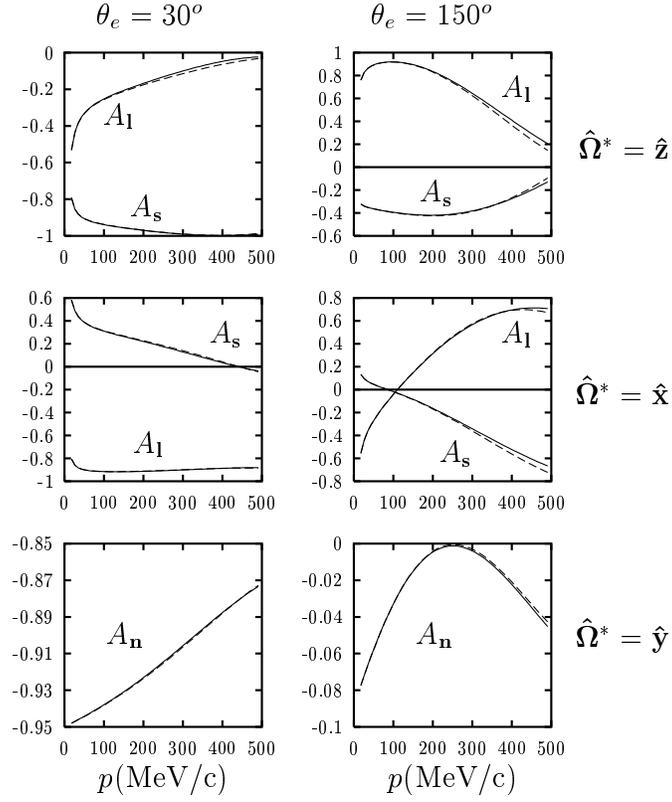}
\end{center}
\caption{Polarization asymmetries $A_{\nl,\ns,\nn}$ as defined 
in~(\ref{asymmetry})
for quasi-perpendicular kinematics. Results are presented for both current
operators, $CC1$ (solid) and $CC2$ (dashed) in the Coulomb gauge.
\label{asymm_perp} 
}
\end{figure}

\begin{figure}
\begin{center}
\rotatebox[origin = lb]{90}{\scalebox{3}[1]}
\includegraphics[scale=0.8,  bb= 80 500 520 750]{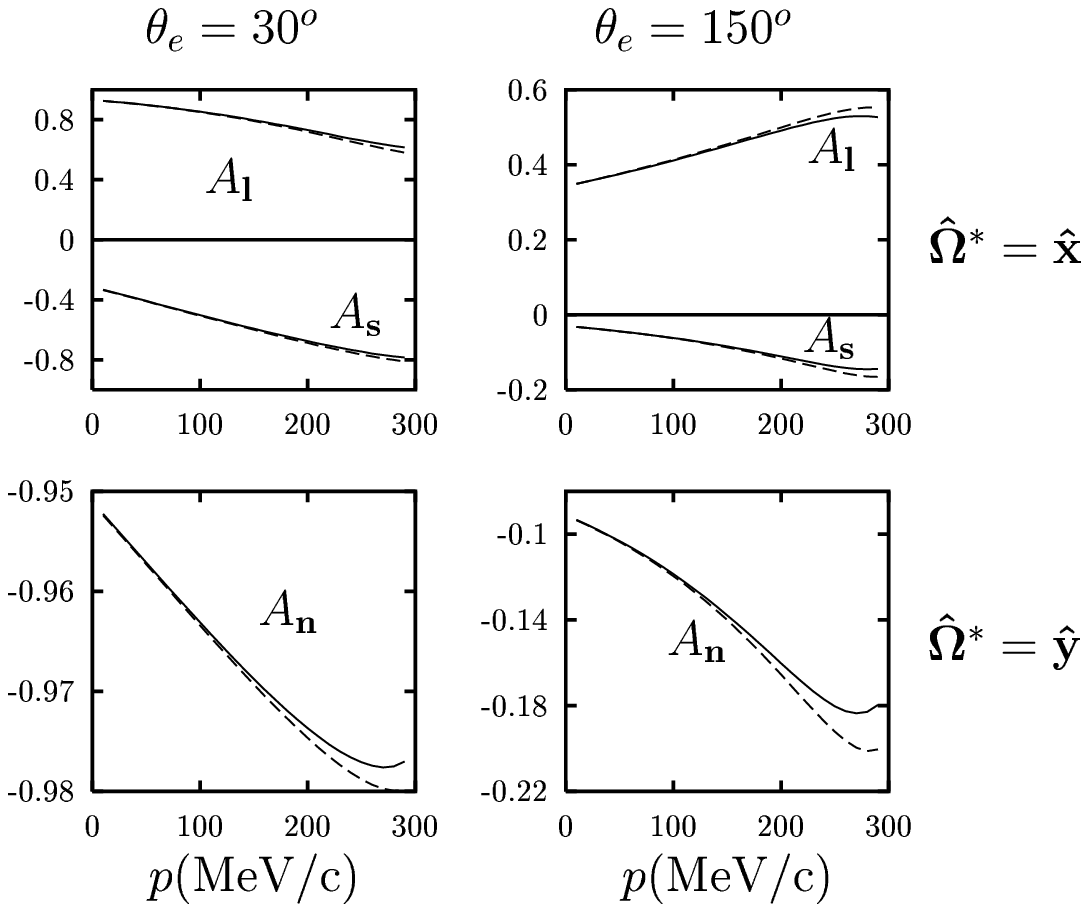}
\end{center}
\caption{Same as Fig.~\ref{asymm_perp} but for parallel kinematics. In
this case the polarized momentum distribution does not enter for the
target oriented along the $z$-axis.\label{asymm_paral}
}
\end{figure}

In Figs.~\ref{asymm_perp} and \ref{asymm_paral} we show the results
for the polarization ratios (\ref{asymmetry}) 
corresponding to quasi-perpendicular and
parallel kinematics, respectively. To simplify the analysis, only the
Coulomb gauge with the two current operator choices has been
considered. Results for the Landau gauge are very similar, whereas
those for the Weyl gauge deviate significantly for missing momentum
values $p\geq 300$ MeV/c.  For both kinematical situations, left
panels correspond to forward electron scattering ($\theta_e=30^o$),
where the main contributions come from the longitudinal responses, and
right panels to backward angles ($\theta_e=150^o$), where the
transverse responses dominate.  In the case of $(q,\omega)$-constant
kinematics (Fig.~\ref{asymm_perp}) and target polarized along $z$ and
$x$ (top and middle panels), only the longitudinal $\nl$ and sideways
$\ns$ polarization asymmetries survive, whereas for the target
orientation along $y$ (bottom panels) only $A_\nn$ differs from
zero. 

\begin{figure}
\begin{center}
\rotatebox[origin = lb]{90}{\scalebox{2}[1]}
\includegraphics[scale=0.8,  bb= 40 340 560 750]{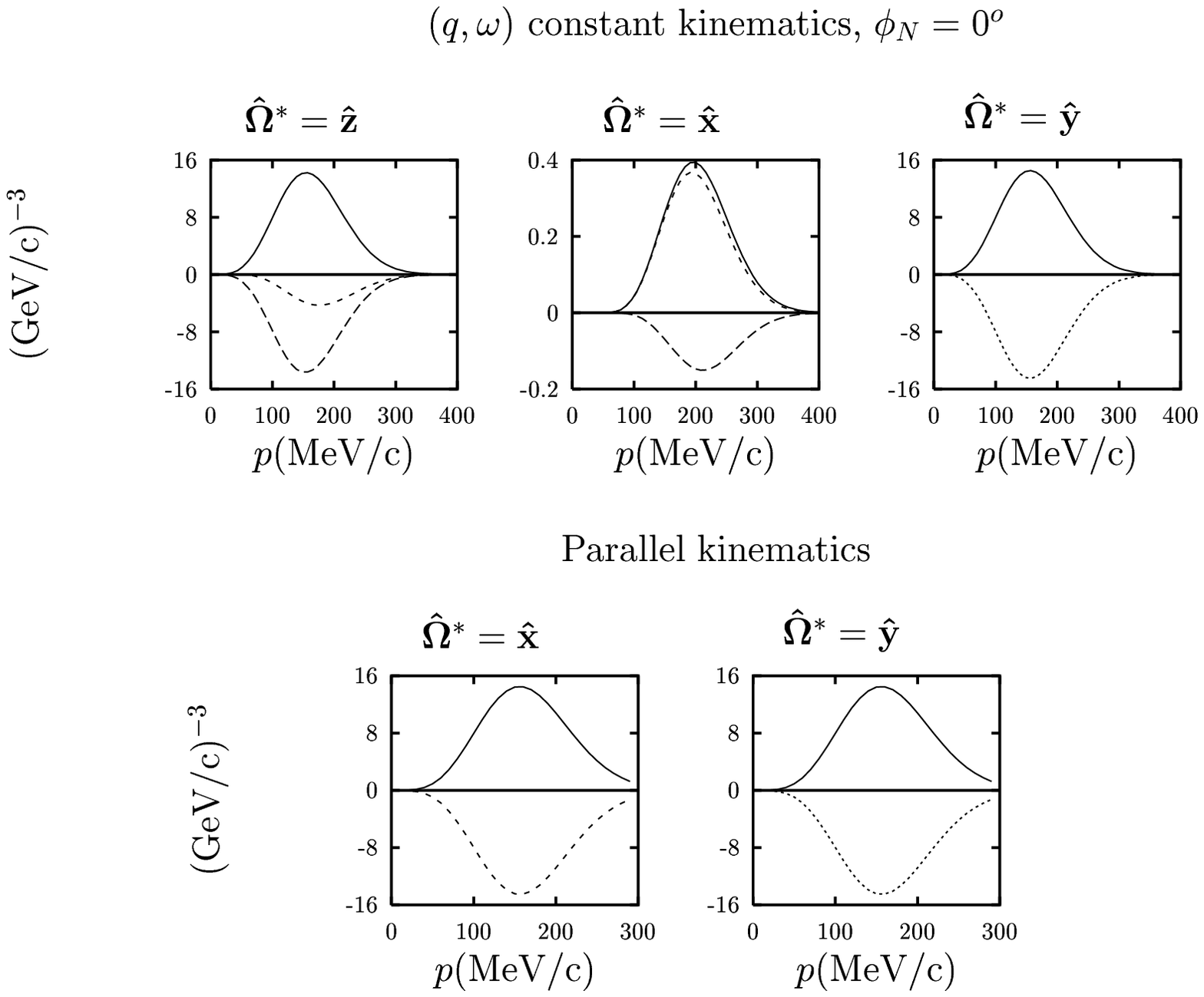}
\end{center}
\caption{Scalar and vector polarized momentum distribution for proton
  knockout from $^{39}$K leading to the residual nucleus in its ground
  state. The labelling of the curves is as follows: $\overline{M}$ (solid),
 $\widehat{M}_x$ (short-dashed), $\widehat{M}_y$ (dotted) and $\widehat{M}_z$ (dashed).
\label{pol_mom}}
\end{figure}

The above results can be easily understood by combining the
single-nucleon responses shown in Figs.~\ref{snresponses_perp_l} and
\ref{snresponses_perp_s} with the general properties of the polarized
momentum distribution, shown in Fig.~\ref{pol_mom} for the case
of interest in this work, i.e., knockout from $^{39}$K.  
The latter is displayed for the two kinematical settings and three 
orientations of the target polarization: along $z$, $x$ and $y$. 
Note that for quasi-perpendicular kinematics (top panels of Fig.~\ref{pol_mom})
the vector term $\widehat{\nM}$ has no $\widehat{M}_y$ contribution for 
$\nOmega^*$
along $z$ and $x$ directions. On the contrary, for $\nOmega^*$
parallel to the $y$-axis, only the vector $\widehat{M}_y$ component
survives and it is equal to $-\overline{M}$. This agrees with the general 
finding, shown in ref.~\cite{Ama02}, that 
$\left|\widehat{\nM}\right|=\left|\overline{M}\right|$ (more details are given
in Appendix C).
In the case of parallel kinematics (bottom panels), the polarized momentum
distribution does not contribute for target orientation along
$z$. Note that in this case $z$ coincides with the missing
momentum direction. It is also important to point out that the scalar
components $\overline{M}$, and consequently the non-vanishing vector
ones, are identical for both target orientations.

Returning to the discussion of the polarization asymmetries, results
in Figs.~\ref{asymm_perp} and~\ref{asymm_paral} show that target and
ejected nucleon polarizations lead in general to very significant
effects in the analysis of $\vec{A}(e,e'\vec{p})$
reactions. Specifically, large polarizations are found for some
directions, reaching almost $\pm 1$, and a clear separation between the
longitudinal ($\nl$) and sideways ($\ns$) nucleon spin directions is
present in all cases.  Concerning Fig.~\ref{asymm_perp}, some specific
symmetries, connected with the results obtained for the single-nucleon
responses shown in previous section, also emerge. This is the
situation for $\theta_e=30^o$, where $A_\ns$ ($A_\nl$) for
$\nOmega^*$ along $z$ is shown to be very similar to $A_\nl$
($-A_\ns$) for $\nOmega^*$ along $x$.  This outcome does not
appear for $\theta_e=150^o$, where the symmetries of the dominant
transverse responses involve the interchange of the $T$ and $TT$ 
components simultaneously
---see~(\ref{sym1}--\ref{sym4}). In the case of target
polarization along $y$ (bottom panels), the only non-vanishing vector
component of the momentum distribution is $\widehat{M}_y$.  
Hence, the numerator in the 
polarization asymmetry arises from the single-nucleon cross section term
$\widehat{\sigma}^{ep}_y$, which only enters for normal ($\nn$) ejected
nucleon polarization.  Because of the general properties of the
polarized momentum distribution, $A_\nn$ does only depend on the
single-nucleon cross section, being
$A_\nn=-\widehat{\sigma}^{ep}_y(\nn)/\overline{\sigma}^{ep}$.  As
shown, the absolute value of $A_\nn$ is close to 1 for
$\theta_e=30^o$, being almost negligible for $\theta_e=150^o$, where
the dominant polarized transverse responses give a much smaller
contribution (see results in previous section).

Next we consider the results for parallel kinematics (Fig.~\ref{asymm_paral}).
Here only the target orientations along $x$ and $y$ directions contribute. In
the former case (upper panel), 
the asymmetries $A_\nl$ and $A_\ns$ show significant effects
and their specific contributions are clearly separated.
This behaviour can be understood from results in section~\ref{sect}
and making use of the general
properties of the polarized momentum distribution~\cite{Cab94,Ama02}. 
Indeed
one simply gets $A_{\nl}=-\widehat{\sigma}^{ep}_x(\nl)/\overline{\sigma}^{ep}$
and $A_{\ns}=-\widehat{\sigma}^{ep}_x(\ns)/\overline{\sigma}^{ep}$ for
$\nOmega^*$ along $x$, and
$A_{\nn}=-\widehat{\sigma}^{ep}_y(\nn)/\overline{\sigma}^{ep}$ for
$\nOmega^*$ in the $y$ direction. As in Fig.~\ref{asymm_perp}, results
corresponding to forward and backward electron scattering angles are very
different, being the global contribution 
bigger in the former situation.

To conclude, we are aware 
of the inherent difficulties in performing double polarized experiments
involving polarized target and ejected nucleon. However, the significant
polarization effects shown in Figs.~\ref{asymm_perp} and \ref{asymm_paral}
are of interest as complementary information to the transferred
polarization ratios $P'_{l,s}$ measured in $A(\vec{e},e'\vec{p})$ processes,
which have been proved to be 
almost insensitive to FSI~\cite{Cris04}.
This is likely  to be also the case for the polarization ratios introduced
in~(\ref{asymmetry}), since they survive in PWIA. 
Moreover, as previously discussed, the ``skewed'' polarization
asymmetries (\ref{asymmetry}) are proved to depend mostly (in some cases
exclusively) on the single-nucleon content of the problem, and this is in
accordance with the behaviour of $P'_{l,s}$ (see~\cite{Cris04}). Therefore,
the analysis of $A_{\nl,\ns,\nn}(\nOmega^*)$ could also contribute to a better
understanding of the properties of nucleons inside nuclei.

\section{Conclusions}

The main objective of this work has been to answer the question: ``does the
simultaneous measurement of target and ejected nucleon polarizations
provide {\em significant} information which could improve our
knowledge of electron scattering reactions?''.  
Double-polarized $\vec{A}(e,e'\vec{p})$ experiments involve technical 
difficulties we are aware of; hence the physical motivations for measuring 
specific observables need to be carefully explored and discussed.  
The present research represents a first step in this direction.

We have focused on the new observables arising 
as polarization transfer from the target to the ejected nucleon.
These are different from the usual polarization transfer observables
measured in $A(\vec{e},e'\vec{p})$ reactions. 
To distinguish from these, we have introduced the term ``skewed''
polarization: it can be pictured as a twist of 
the spin of the hit nucleon inside the polarized target.

The analysis of ``skewed'' polarization in $\vec{A}(e,e'\vec{p})$ has been 
carried out within the framework of PWIA. 
Obviously, PWIA is too simple to provide realistic cross sections and 
response functions to be compared with the experiment. 
However, ratios of cross sections are expected to be much 
less sensitive to FSI effects than the cross sections themselves.
This has been shown to be the case for the transferred polarization 
asymmetries $P'_{l,s}$, where PWIA predictions for missing momenta up to 
the Fermi momentum coincide with much more elaborated calculations involving
relativistic and non-relativistic distorted wave 
approaches~\cite{Kaz04,Cris04}.
Therefore, we are reasonably confident that our predictions for the 
polarization asymmetries introduced in~(\ref{asymmetry}) 
will be roughly valid within more sophisticated approximations. 

A further merit of PWIA lies on the fact that, besides its simplicity, 
it allows to treat
relativistic aspects of the reaction in a complete way.  Moreover,
polarization dependence in the general scattering problem also emerges
within PWIA in a very clear way. So, the analysis presented in Section
\ref{sec:formalism} contains some of the basic findings of this work.  
Specifically, the expressions given in~(\ref{resp1},\ref{resp2}) allow one to
identify how spin degrees of freedom get organized in the different
observables. A crucial difference comes out between the
electron-polarized and electron-unpolarized responses.  In the former,
the whole dependence upon the spin goes simply as the sum of the bound 
and ejected nucleon four-spins, whereas for the latter, the product of
four-spins is involved. Obviously, the scalar and vector momentum
distributions, both dependent in general on the target polarization,
must be also added in order to get final observables.

In presenting results for the new quantities needed to describe the 
process, we have first analyzed the ``skewing'' single-nucleon response functions 
for different off-shell prescriptions of the electromagnetic current. 
Second, we have presented predictions for the polarization asymmetries
associated to a polarized $^{39}$K target.
The preliminary study performed in the
present work indicates that experiments with simultaneously polarized
target and recoil nucleon can provide additional information to that
obtained from the separate investigation of $\vec{A}(\vec{e},e'p)$ and
$A(\vec{e},e'\vec{p})$ processes.  The predictions presented here for
the asymmetries $A_{\nl,\ns,\nn}(\nOmega^*)$, setting the scale of values 
expected for these new observables, constitute a promising starting point. 
Moreover, the strong dependence of the above asymmetries on the
single-nucleon ingredients makes these observables suited to
investigate nucleon properties, complementing the
analysis of transferred polarization ratios $P'_{l,s}$.

\subsection*{Acknowledgements}

This work was partially supported by funds provided by DGI (Spain) and FEDER
funds, under Contracts Nos. BFM2002-03218, BFM2002-03315 and
FPA2002-04181-C04-04 and by the Junta de Andaluc\'{\i}a, and by the INFN-CICYT
collaboration agreement (project ``Study of relativistic dynamics in electron
and neutrino scattering'').

\section*{Appendix A}

We summarize the results for the single-nucleon tensors. For the two off-shell
vertices we find the following. For the CC1 case:

\ba
&&4M_N^2{\cal S}^{\mu\nu}(\xi,S_N)=G_M^2\left[(1-\xi\cdot S_N)
(\overline{P}^\mu P_N^\nu+\overline{P}^\nu P_N^\mu)+
\frac{\overline{Q}^2}{2}g^{\mu\nu}(1-\xi\cdot S_N) \right. \nonumber \\
 &+&\left.
P_N\cdot\xi(\overline{P}^\mu S_N^\nu+\overline{P}^\nu S_N^\mu)+
\overline{P}\cdot S_N(P_N^\mu\xi^\nu+P_N^\nu\xi^\mu-g^{\mu\nu}P_N\cdot\xi)
+\frac{\overline{Q}^2}{2}(\xi^\mu S_N^\nu+\xi^\nu S_N^\mu)\right] \nonumber \\
&-&\frac{F_2}{2}G_M\left[\overline{P}\cdot S_N (\xi^\mu R^\nu+\xi^\nu R^\mu)+
P_N\cdot\xi (S_N^\mu R^\nu+S_N^\nu R^\mu)+2(1-\xi\cdot S_N)R^\mu R^\nu\right]
\nonumber \\
&+&\frac{F_2^2}{4M_N^2}R^\mu R^\nu\left[\overline{P}\cdot S_N P_N\cdot \xi +
\left(2M_N^2-\frac{\overline{Q}^2}{2}\right)(1-\xi\cdot S_N)\right]
\, ,
\ea

\ba
&&4M_N^2{\cal A}^{\mu\nu}(\xi,S_N)= M_N^2G_M^2\epsilon^{\mu\nu\sigma\alpha}
(\xi+S_N)_\sigma\overline{Q}_\alpha  \nonumber \\
&+& \frac{F_2}{2M_N}G_M(R^\mu\epsilon^{\sigma\alpha\beta\nu}-
R^\nu\epsilon^{\sigma\alpha\beta\mu})(\xi+S_N)_\sigma \overline{P}_\alpha P_{N\beta} \, ,
\ea
where $R^\mu\equiv(\overline{P}+P_N)^\mu$ and 
$\overline{Q}^\mu\equiv(P_N-\overline{P})^\mu$.

For the CC2 case, we have

\ba
&&4M_N^2{\cal S}^{\mu\nu}(\xi,S_N)=
F_1^2\left[(1-\xi\cdot S_N)\left(\overline{P}^\mu P_N^\nu+\overline{P}^\nu P_N^\mu+
\frac{\overline{Q}^2}{2}g^{\mu\nu}\right)+\frac{\overline{Q}^2}{2}
(\xi^\mu S_N^\nu+\xi^\nu S_N^\mu) \right. \nonumber \\
&+& \left. P_N\cdot \xi (\overline{P}^\mu S_N^\nu+\overline{P}^\nu S_N^\mu)+
\overline{P}\cdot S_N(P_N^\mu\xi^\nu+P_N^\nu\xi^\mu)-g^{\mu\nu}P_N\cdot\xi \overline{P}\cdot S_N)
\right] \nonumber \\
&+& \frac{F_1F_2}{2}\left[2Q\cdot \overline{Q}(\xi^\mu S_N^\nu+\xi^\nu S_N^\mu)-
(1-\xi\cdot S_N)(\overline{Q}^\mu Q^\nu+\overline{Q}^\nu Q^\mu-2g^{\mu\nu}Q\cdot\overline{Q})
\right. \nonumber \\
&+& \left. 2Q\cdot\xi (\overline{P}^\mu S_N^\nu+\overline{P}^\nu S_N^\mu
-g^{\mu\nu}\overline{P}\cdot S_N)-
2Q\cdot S_N (P_N^\mu\xi^\nu+P_N^\nu\xi^\mu
-g^{\mu\nu}P_N\cdot\xi) \right. \nonumber \\
&+& \left. \overline{P}\cdot S_N (\xi^\mu Q^\nu+\xi^\nu Q^\mu)-
P_N\cdot\xi (S_N^\mu Q^\nu+S_N^\nu Q^\mu)\right] \nonumber \\
&+& \frac{F_2^2}{4M_N^2}\left[\left( Q\cdot S_N P_N\cdot\xi+Q\cdot P_N(1-\xi\cdot S_N)\right)
(\overline{P}^\mu Q^\nu+\overline{P}^\nu Q^\mu) \right. \nonumber \\
&+& \left. \left(\overline{P}\cdot S_N Q\cdot\xi+\overline{P}\cdot Q(1-\xi\cdot S_N)\right)
(P_N^\mu Q^\nu+P_N^\nu Q^\mu) \right. \nonumber \\
&-& \left. \left(Q^2(1-\xi\cdot S_N)+2Q\cdot\xi Q\cdot S_N\right)
(\overline{P}^\mu P_N^\nu+\overline{P}^\nu P_N^\mu) \right. \nonumber \\
&-& \left. (Q^2 P_N\cdot\xi-2Q\cdot\xi Q\cdot P_N)(\overline{P}^\mu S_N^\nu+\overline{P}^\nu S_N^\mu)
\right. \nonumber \\
&-& \left. (Q^2\overline{P}\cdot S_N-2Q\cdot P_N Q\cdot S_N)(P_N^\mu\xi^\nu+P_N^\nu\xi^\mu)
\right. \nonumber \\
&+& \left. \left(Q^2(2M_N^2-\frac{\overline{Q}^2}{2})-2Q\cdot P_N Q\cdot\overline{P}\right)
(\xi^\mu S_N^\nu+\xi^\nu S_N^\mu-g^{\mu\nu}\xi\cdot S_N) \right. \nonumber \\
&-& \left. \left(Q\cdot S_N(2M_N^2-\frac{\overline{Q}^2}{2})-\overline{P}\cdot S_N Q\cdot P_N\right)
(\xi^\mu Q^\nu+\xi^\nu Q^\mu-g^{\mu\nu}Q\cdot\xi) \right. \nonumber \\
&-& \left. \left(Q\cdot\xi(2M_N^2-\frac{\overline{Q}^2}{2})-P_N\cdot\xi Q\cdot\overline{P}\right)
(S_N^\mu Q^\nu+S_N^\nu Q^\mu-g^{\mu\nu}Q\cdot S_N) \right. \nonumber \\
&-& \left. \left((1-\xi\cdot S_N)(2M_N^2-\frac{\overline{Q}^2}{2})+P_N\cdot\xi \overline{P}\cdot S_N
\right)Q^\mu Q^\nu \right. \nonumber \\
&+& \left.  g^{\mu\nu}\left(Q^2(2M_N^2-\frac{\overline{Q}^2}{2}+\overline{P}\cdot S_N
P_N\cdot \xi) \right. \right. \nonumber \\
&-& \left. \left. Q\cdot P_N(Q\cdot\overline{P}+Q\cdot\xi\overline{P}\cdot S_N)-
 Q\cdot\overline{P}(Q\cdot P_N+Q\cdot S_N P_N\cdot\xi)\right)\right] \, ,
\ea
where $Q^\mu\equiv K^\mu-K^{'\mu}$ is the four-momentum transferred 
in the process.
Finally
\ba
&&4M_N^2{\cal A}^{\mu\nu}(\xi,S_N)=M_NF_1^2\epsilon^{\alpha\beta\mu\nu}(\xi+S_N)_\alpha
\overline{Q}_\beta \nonumber \\
&+& \frac{F_1F_2}{2M_N}\left[\epsilon^{\alpha\beta\mu\nu}(
    2P_N\cdot Q\overline{P}_\alpha\xi_\beta+2\overline{P}\cdot Q P_{N\alpha}S_{N\beta})
\right. \nonumber \\
&+&\left. 2M_N^2\epsilon^{\alpha\beta\mu\nu}(\xi+S_N)_\alpha Q_\beta \right. \nonumber \\
&+& \left. (Q^\mu\epsilon^{\alpha\beta\gamma\nu}-Q^\nu\epsilon^{\alpha\beta\gamma\mu})
(S_N-\xi)_\alpha P_{N\beta}\overline{P}_\gamma \right] \nonumber \\
&+& \frac{F_2^2}{4M_N}\left[
  \epsilon^{\alpha\beta\mu\nu}(2P_N\cdot Q\xi_\alpha Q_\beta-2\overline{P}\cdot Q S_{N\alpha}
Q_\beta)+Q^2\epsilon^{\alpha\beta\mu\nu}(\overline{P}+P_N)_\alpha(\xi-S_N)_\beta
\right. \nonumber \\
&+&\left. (Q^\mu\epsilon^{\alpha\beta\gamma\nu}-Q^\nu\epsilon^{\alpha\beta\gamma\mu})
(\xi-S_N)_\alpha(\overline{P}+P_N)_\beta Q_\gamma \right] \, .
\ea

\section*{Appendix B}

In this appendix we show the explicit expressions of the scalar and vector 
single-nucleon responses by taking the appropriate components of the general tensor
${\cal W}^{\mu\nu}(\theta_R,\phi_R;S_N)$ as introduced in~(\ref{sntensor2}). To make clear
the analysis, we fix the following notation:
$\widehat{\ncalR}^{\alpha(\alpha')}\equiv\left(\widehat{\cal R}^{\alpha(\alpha')}_x,
\widehat{\cal R}^{\alpha(\alpha')}_y,\widehat{\cal R}^{\alpha(\alpha')}_z \right)$ 
for the three vector components, and $\overline{\cal R}^{\alpha(\alpha')}$
the scalar one. In terms of the tensor components given by the
Lorentz indices $\mu,\nu$, referred to the $1-2-3$ hadron-plane-system of axes, the
final expressions result:
\begin{itemize}
\item {\sl Electron-unpolarized responses}
\ba
&&\overline{\cal R}^L={\cal S}^{00} \nonumber \\
&&\widehat{\cal R}^L_z={\cal S'}^{00}(0,0,S_N) \nonumber \\
&&\widehat{\cal R}^L_x={\cal S'}^{00}(\frac{\pi}{2},0,S_N) \nonumber \\
&&\widehat{\cal R}^L_y={\cal S'}^{00}(\frac{\pi}{2},\frac{\pi}{2},S_N) \, .
\ea
\ba
&&\overline{\cal R}^T={\cal S}^{11}+{\cal S}^{22} \nonumber \\
&&\widehat{\cal R}^T_z={\cal S'}^{11}(0,0,S_N)+{\cal S'}^{22}(0,0,S_N) \nonumber \\
&&\widehat{\cal R}^T_x={\cal S'}^{11}(\pi/2,0,S_N)+{\cal S'}^{22}(\pi/2,0,S_N) \nonumber \\
&&\widehat{\cal R}^T_y={\cal S'}^{11}(\pi/2,\pi/2,S_N)+{\cal S'}^{22}(\pi/2,\pi/2,S_N) \, .
\ea
\ba
&&\overline{\cal R}^{TT}=\left({\cal S}^{22}-{\cal S}^{11}\right)\cos 2\phi_N \nonumber \\
&&\widehat{\cal R}^{TT}_z=\left[{\cal S'}^{22}(0,0,S_N)-{\cal S'}^{11}(0,0,S_N)\right]
\cos 2\phi_N+2{\cal S'}^{12}(0,0,S_N) \sin 2\phi_N
 \nonumber \\
&&\widehat{\cal R}^{TT}_x=\left[{\cal S'}^{22}(\pi/2,0,S_N)-{\cal S'}^{11}(\pi/2,0,S_N)
\right]\cos 2\phi_N+2{\cal S'}^{12}(\pi/2,0,S_N)\sin 2\phi_N
 \nonumber \\
&&\widehat{\cal R}^{TT}_y=\left[
{\cal S'}^{22}(\pi/2,\pi/2,S_N)-{\cal S'}^{11}(\pi/2,\pi/2,S_N)\right]\cos 2\phi_N+
2{\cal S'}^{12}(\pi/2,\pi/2,S_N)\sin 2\phi_N
\, . \nonumber \\
&&
\ea
\ba
&&\overline{\cal R}^{TL}=2\sqrt{2}{\cal S}^{01}\cos\phi_N \nonumber \\
&&\widehat{\cal R}^{TL}_z=2\sqrt{2}\left[{\cal S'}^{01}(0,0,S_N)\cos\phi_N-
{\cal S'}^{02}(0,0,S_N)\sin\phi_N\right] \nonumber \\
&&\widehat{\cal R}^{TL}_x=2\sqrt{2}\left[{\cal S'}^{01}(\pi/2,0,S_N)\cos\phi_N-
{\cal S'}^{02}(\pi/2,0,S_N)\sin\phi_N\right] \nonumber \\
&&\widehat{\cal R}^{TL}_y=2\sqrt{2}\left[{\cal S'}^{01}(\pi/2,\pi/2,S_N)\cos\phi_N-
{\cal S'}^{02}(\pi/2,\pi/2,S_N)\sin\phi_N\right] \, .
\ea
\item {\sl Electron-polarized responses}
\ba
&&\overline{\cal R}^{T'}=-2{\cal A}^{12}(S_N) \nonumber \\
&&\widehat{\cal R}^{T'}_z=-2{\cal A'}^{12}(0,0) \nonumber \\
&&\widehat{\cal R}^{T'}_x=-2{\cal A'}^{12}(\pi/2,0) \nonumber \\
&&\widehat{\cal R}^{T'}_y=-2{\cal A'}^{12}(\pi/2,\pi/2) \, .
\ea
\ba
&&\overline{\cal R}^{TL'}=-2\sqrt{2}\left[{\cal A}^{01}(S_N)\sin\phi_N+{\cal A}^{02}(S_N)\cos\phi_N\right]
\nonumber \\
&&\widehat{\cal R}^{TL'}_z=-2\sqrt{2}\left[{\cal A'}^{01}(0,0)\sin\phi_N+
{\cal A'}^{02}(0,0)\cos\phi_N\right]
\nonumber \\
&&\widehat{\cal R}^{TL'}_x=-2\sqrt{2}\left[{\cal A'}^{01}(\pi/2,0)\sin\phi_N+
{\cal A'}^{02}(\pi/2,0)\cos\phi_N\right]
\nonumber \\
&&\widehat{\cal R}^{TL'}_y=-2\sqrt{2}\left[{\cal A'}^{01}(\pi/2,\pi/2)\sin\phi_N+
{\cal A'}^{02}(\pi/2,\pi/2)\cos\phi_N\right] \, .
\ea
\end{itemize}

\section*{Appendix C}

In this Appendix we discuss in detail the geometrical interpretation of the
polarized momentum distribution. In the case of interest in this work, 
a proton knocked out from the $d_{3/2}$ shell of $^{39}$K and the daughter nucleus in its
ground state, $J_B=0$, the vector field $\nchi$ introduced in (\ref{spin}) reduces to
\begin{equation}
\nchi = 2\frac{\nOmega^*\cdot\np}{p^2} \np -\nOmega^* \, .
\end{equation}
The symmetries of the momentum distribution can then be explained
by taking specific directions of the target polarization $\nOmega^*$:
\begin{itemize}
\item $\nOmega^*=\hat\nz$. In this case the nucleus is polarized along $\nq$. 
For quasi-elastic kinematics $p_N\simeq q$, hence we can write
$\cos\theta \simeq -\frac{p}{2q}$, where $\theta$ is the angle between 
$\np$ and $\nq$. Therefore
\begin{equation}
\nchi=-\frac{\np+\nq}{q}\,.
\end{equation}
For low values of the missing momentum (compared with $q$) we can also write 
$\sin\theta\simeq 1$ to leading order in $p/q$.
The components of $\nchi$ are then
\begin{equation}
\chi_x = -\frac{p_x}{q}\simeq -\frac{p}{q}, \kern 1cm
\chi_y = 0, \kern 1cm
\chi_z = -\frac{p_z}{q}-1 \simeq -1 \,.
\end{equation}
Therefore the following relations between vector and scalar
momentum distributions
\begin{equation}
\widehat{M}_x \simeq -\left(\frac{p}{q}\right) \overline{M},
\kern 1cm
\widehat{M}_y =0,
\kern 1cm
\widehat{M}_z \simeq - \overline{M},
\end{equation}
hold, in accord with the results of the upper-left panel of Fig.~\ref{pol_mom}.
\item $\nOmega^*=\hat\nx$. In this case $\np\cdot\nOmega^*=p_x\simeq p$ and 
the field $\nchi$ is
\begin{equation}
\nchi \simeq 2\frac{\np}{p}-\hat\nx.
\end{equation}
Hence we find to first order in $p/q$ 
\begin{equation}
\chi_x=2\frac{p_x}{p}-1 \simeq 1,
\kern 1cm
\chi_y =0,
\kern 1cm
\chi_z = 2\frac{p_z}{p}\simeq -\frac{p}{q}
\end{equation}
and for the momentum distribution components
\begin{equation}
\widehat{M}_x \simeq \overline{M},
\kern 1cm
\widehat{M}_y =0,
\kern 1cm
\widehat{M}_z \simeq - \frac{p}{q}\overline{M},
\end{equation}
explaining the results of the upper-middle panel of  Fig.~\ref{pol_mom}.
\item $\nOmega^*=\hat\ny$. In this case $\np$ is perpendicular to $\nOmega^*$,
so $\nchi=-\hat\ny$. Hence
\begin{equation}
\widehat{M}_x =
\widehat{M}_z =0,
\kern 1cm
\widehat{M}_y = -\overline{M},
\end{equation}
giving the results of the 
upper-right panel of  Fig.~\ref{pol_mom}.
\end{itemize} 
A similar analysis can be applied to the results for parallel kinematics
(bottom panels of Fig.~\ref{pol_mom}).


\begin{thebibliography}{10}


\bibitem{Frul85} S. Frullani and J. Mougey, Adv. Nucl. Phys. {\bf 14} (1985).

\bibitem{RaDo89} A.S. Raskin, T.W. Donnelly,
                 Ann. Phys. {\bf 191} (1989) 78.

\bibitem{Kel96}  J. J. Kelly, Adv. Nucl. Phys. {\bf 23} (1996) 75.

\bibitem{Bof96}  S. Boffi, C. Giusti, F. D. Pacati, M. Radici,
                 {\em Electromagnetic response of atomic nuclei},
                 Oxford University Press (1996).

\bibitem{Donnelly:1991qy}
T.~W.~Donnelly, M.~J.~Musolf, W.~M.~Alberico, M.~B.~Barbaro, A.~De Pace and A.~Molinari,
Nucl.\ Phys.\ A {\bf 541} (1992) 525.

\bibitem{Musolf:1993tb} M.~J.~Musolf, T.~W.~Donnelly, J.~Dubach, S.~J.~.~Pollock, S.~Kowalski and E.~J.~Beise,
Phys.\ Rept.\  {\bf 239} (1994) 1.

\bibitem{Alberico:1993ur}
W.~M.~Alberico, M.~B.~Barbaro, A.~De Pace, T.~W.~Donnelly and A.~Molinari,
Nucl.\ Phys.\ A {\bf 563} (1993) 605.

\bibitem{Barbaro:1993jg}
M.~B.~Barbaro, A.~De Pace, T.~W.~Donnelly and A.~Molinari,
Nucl.\ Phys.\ A {\bf 569} (1994) 701.

\bibitem{Barbaro:1995ez}
M.~B.~Barbaro, A.~De Pace, T.~W.~Donnelly and A.~Molinari,
Nucl.\ Phys.\ A {\bf 596} (1996) 553.

\bibitem{Barbaro:1995gp}
M.~B.~Barbaro, A.~De Pace, T.~W.~Donnelly and A.~Molinari,
Nucl.\ Phys.\ A {\bf 598} (1996) 503.

\bibitem{Woo98} R.J. Woo et al.,
                Phys. Rev. Lett. {\bf 80} (1998) 456.

\bibitem{TJNAF} TJNAF proposal 89-033, C. Glashausser contact person.

\bibitem{Udi00} J.M. Ud\'{\i}as, J.R. Vignote,
                 Phys. Rev. {\bf C62} (2000) 034302.

\bibitem{Ito97}  H. Ito, S.E. Koonin, R. Seki, 
                 Phys. Rev. C {\bf 56} 3231 (1997).

\bibitem{Ryc99}  J. Ryckebusch, D. Debruyne, W. Van Nespen, and S. Janssen,
                 Phys. Rev. C {\bf 60} 034604 (1999).

\bibitem{Kel99}  J.J. Kelly, 
                 Phys. Rev. C {\bf 59} 3256 (1999).

\bibitem{Kel99b} J.J. Kelly, 
                 Phys. Rev. C {\bf 60} 044609 (1999).

\bibitem{Joh99}  J.I. Johanson and H.S. Sherif,
                 Phys. Rev. C {\bf 59} 3481 (1999).

\bibitem{Kaz04} F. Kazemi Tabatabaei, J.E. Amaro, J.A. Caballero,
                 Phys. Rev. {\bf C69} (2004) 064607.

\bibitem{Mal99} S. Malov, 
                PhD thesis, New Brunswick, New Jersey, (1999), unpublished.

\bibitem{Mal00} S. Malov et al., 
                Phys. Rev. {\bf C62} (2000) 057302.

\bibitem{Die01}
S. Dieterich et al.,
Phys. Lett. {\bf B500} (2001) 47.

\bibitem{Strauch03}
S. Strauch et al.,
Phys. Rev. Lett. {\bf 91} (2003) 052301.

\bibitem{Cris04} M.C. Mart\'{\i}nez, J.R. Vignote, J.A. Caballero, 
                 T.W. Donnelly, E. Moya de Guerra, J.M. Ud\'{\i}as,
                 Phys. Rev. {\bf C69} (2004) 034604.

\bibitem{Boffi88}
S. Boffi, C. Giusti, F.D. Pacati,
Nucl. Phys. {\bf A476} (1988) 617.

\bibitem{Cab93}
J.A. Caballero, T.W. Donnelly, G.I. Poulis,
Nucl. Phys. {\bf A555} (1993) 709.

\bibitem{Cab94}
J.A. Caballero, T.W. Donnelly, G.I. Poulis, E. Garrido, E. Moya de Guerra,
Nucl. Phys. {\bf A577} (1994) 528.

\bibitem{Edu95}
E. Garrido, J.A. Caballero, E. Moya de Guerra, P. Sarriguren, J.M. Ud\'{\i}as,
Nucl. Phys. {\bf A584} (1995) 256.

\bibitem{JAC95}
J.A. Caballero, E. Garrido, E. Moya de Guerra, P. Sarriguren, J.M. Ud\'{\i}as,
Ann. Phys. {\bf 239} (1995) 351.

\bibitem{Ama98b} J.E. Amaro and T.W. Donnelly,
                 Ann. Phys. (N.Y.) {\bf 263} 56 (1998).

\bibitem{Ama99}
J.E. Amaro, T.W. Donnelly,
Nucl. Phys. {\bf A 646} (1999) 187.

\bibitem{Ama02}
J.E. Amaro, T.W. Donnelly,
Nucl. Phys. {\bf A 703} (2002) 541.

\bibitem{Are04} H. Arenh\"ovel, W. Leidemann and E.L. Tomusiak,
                nucl-th/0407053.

\bibitem{Are02} H. Arenh\"ovel, W. Leidemann and E.L. Tomusiak,
                Eur. Phys. Jou. {\bf A} (2002) 491.

\bibitem{Are00} H. Arenh\"ovel, W. Leidemann and E.L. Tomusiak,
                Few Body Syst. 28 (2000) 147.

\bibitem{Are98} H. Arenh\"ovel, W. Leidemann and E.L. Tomusiak,
                Nucl. Phys. {\bf A641} (1998) 517.
 
\bibitem{Are95} H. Arenh\"ovel, W. Leidemann and E.L. Tomusiak,
                Phys. Rev. C 52 (1995) 1232.

\bibitem{Are92} H. Arenh\"ovel, W. Leidemann and E.L. Tomusiak,
                Phys. Rev.  C 46 (1992) 455.

\bibitem{Ama04} J.E. Amaro, J.A. Caballero, T.W. Donnelly, 
                F. Kazemi Tabatabaei, in preparation.

\bibitem{Kaz03} F. Kazemi Tabatabaei, J.E. Amaro, J.A. Caballero,
                 Phys. Rev. {\bf C68} (2003) 034611.

\bibitem{Cab98a}
J.A. Caballero, T.W. Donnelly, E. Moya de Guerra, J.M. Ud\'{\i}as,
Nucl. Phys. {\bf A632} (1998) 323.

\bibitem{Cab98b}
J.A. Caballero, T.W. Donnelly, E. Moya de Guerra, J.M. Ud\'{\i}as,
Nucl. Phys. {\bf A643} (1998) 189.

\bibitem{Cris02}
M.C. Mart\'{\i}nez, J.A. Caballero, T.W. Donnelly,
Nucl. Phys. {\bf A 707} (2002) 83; {\bf A 707} (2002) 121. 

\bibitem{Vign04} J.R. Vignote, M.C. Mart\'{\i}nez, J.A. Caballero, 
E. Moya de Guerra, J.M. Ud\'{\i}as,
Phys. Rev. {\bf C70} (2004) 044608.

\bibitem{For83}
T. de Forest,
Nucl. Phys. {\bf A392} (1983) 232.

\bibitem{notes}
J.E. Amaro, M.B. Barbaro, J.A. Caballero,  in preparation.

\bibitem{Ama96}
J.E. Amaro, J.A. Caballero, T.W. Donnelly, E. Moya de Guerra,  
Nucl. Phys. {\bf A 611} (1996) 163.

\end{thebibliography}


\end{document}